\documentclass{aa}
\usepackage{txfonts,graphicx,multirow}
\usepackage{natbib,color} 
\bibpunct{(}{)}{;}{a}{}{,} 

\title{The Chromospheric Telescope} 
\subtitle{}

\author{C. Bethge\inst{\ref{KIS},}\inst{\ref{HAO}}
\and H. Peter\inst{\ref{MPS},}\inst{\ref{KIS}}  
\and T. J. Kentischer\inst{\ref{KIS}}
\and C. Halbgewachs\inst{\ref{KIS}}
\and D. F. Elmore\inst{\ref{NSO}}
\and C. Beck\inst{\ref{IAC}}
}

\institute{Kiepenheuer-Institut f\"ur Sonnenphysik, Sch\"oneckstr. 6, 79104 Freiburg, Germany\label{KIS}
\and High Altitude Observatory, National Center for Atmospheric
Research\thanks{The National Center for Atmospheric Research is
  sponsored by the National Science Foundation.}, P.O. Box 3000, Boulder, CO 80307, USA\newline\email{bethge@ucar.edu}\label{HAO}
\and Max-Planck-Institut f\"ur Sonnensystemforschung, 37191 Katlenburg-Lindau, Germany\label{MPS}
\and National Solar Observatory, 3010 Coronal Loop, Sunspot, NM 88349, USA\label{NSO}
\and Instituto de Astrof{\'\i}sica de Canarias (CSIC), Via Lactea, E-38205 La Laguna, Tenerife, Spain\label{IAC}}

\date{Received 9 June 2011 / Accepted 27 July 2011}

\abstract{}
  {We introduce the \emph{Chro}mospheric \emph{Tel}escope (ChroTel) at the
  Observatorio del Teide in Iza$\rm{\tilde{n}}$a on Tenerife as a new multi-wavelength
  imaging telescope for full-disk synoptic observations of the solar
  chromosphere. We describe the design of the instrument and summarize
  its performance during the first one and a half years of
  operation. We present a method to derive line-of-sight velocity maps of the
  full solar disk from filtergrams taken
  in and near the \mbox{\ion{He}{i}} infrared line at 10830\,\AA.}
  {ChroTel observations are conducted using Lyot-type filters for the chromospheric
  lines of \ion{Ca}{ii} K, H$\alpha$, and \mbox{\ion{He}{i}}
  10830\,\AA. The instrument operates 
  autonomically and gathers imaging data in all three channels with a
  cadence of down to one minute. The use of a tunable filter for the
  \mbox{\ion{He}{i}} line allows us to determine line-shifts by calibrating the line-of-sight velocity 
  maps derived from the filtergram intensities with spectrographic
  data from the Tenerife Infrared Polarimeter at high spatial and spectral resolution.} 
  {The robotic operation and automated data reduction have proven to
    operate reliably in the first one and and half years. The
    achieved spatial resolution of the data is close to the
    theoretical limit of 2 arcsec in H$\alpha$ and \mbox{\ion{Ca}{ii} K}
    and 3 arcsec in \mbox{\ion{He}{i}}. Line-of-sight velocities in
    \mbox{\ion{He}{i}} can be determined
    with a precision of better than 3-4~km\,s$^{-1}$ when co-temporal
    spectrographic maps are available for calibration.}
    {ChroTel offers
    a unique combination of imaging in the most important
    chromospheric lines, along with the possibility to determine
    line-of-sight velocities in one of the lines. This is of interest for scientific
    investigations of large-scale structures in the solar chromosphere, as well as for
    context imaging of high-resolution solar observations.}

\keywords{{Sun: chromosphere} - {Telescopes}}

\begin{document}
\maketitle
\section{Introduction}
Full-disk observations of the Sun, from both the ground and
space, have always played an important role in solar physics because they
are not only essential to global solar studies, e.g., for helioseismology
and solar activity, but also provide contextual information for
observations at high resolution that study only small parts of the
solar disk. Among the most prominent examples are the GONG network
\citep{Harvey1996} and the MDI instrument onboard SOHO
\citep{Scherrer1995}. It is widely believed that key insight
emerges predominantly from high-resolution observations advancing to
smaller and smaller structures in the Sun's atmosphere. Ground-based full-disk
instruments commonly share the disadvantage of a comparably low
spatial resolution. They must be operated seeing-limited because
adaptive optics can correct only for a small portion of the solar
disk. However, despite this disadvantage, they are still vital in solar
research - first and foremost for structures on larger scales up to
the complete disk, but also for rarely occurring events that are often
missed when examining only a small fraction of the disk. Here, we
present such an instrument for observations of the
solar chromosphere. 

Chromospheric full-disk observations are often performed in the
spectral lines of \mbox{\ion{Ca}{ii} K} (3933.7~\AA) and H$\alpha$
(6562.8~\AA) because both lines are known to indicate the presence
and the structure of magnetic fields. Prominent examples include
the RISE-PSPT instruments observing in \mbox{\ion{Ca}{ii} K}
\citep{Ermolli1998} at the Osservatorio Astronomico di Roma (OAR) and
the Mauna Loa Solar Observatory (MLSO), as well as the H$\alpha$
patrol instrument of the Kanzelh\"ohe Observatory \citep{Otruba1999}
in Austria, which is part of the global H$\alpha$ network
\citep{Steinegger2000}. 

The \mbox{\ion{He}{i}} triplet at 10830~\AA$\;$has become
an increasingly popular diagnostic of the chromosphere, although its
formation process is complicated. In spectropolarimetry, this is due
to the sensitivity of the polarization signal to both the Hanle and the
Zeeman effect, rendering the line(s) an ideal tool for both strong and
weak magnetic field regions \citep[e.g.,][]{Rueedi1995,Trujillo2005}. 
However, it is also very useful for full-disk observations because the line is formed entirely
in the upper chromosphere (that is at chromospheric temperatures) or
the lower transition region with no photospheric contributions. In addition, the
line formation is strongly influenced by coronal emission because the
triplet states are mainly populated by the
photoionization-recombination mechanism \citep{Centeno2008}. This
permits an estimation of coronal activity in ground-based observations
- coronal holes for example are clearly visible in \mbox{\ion{He}{i}}
10830~\AA~\citep[see e.g.,][]{Dupree1996}. However,
  observations can still suffer from photospheric contamination in
  regions with little photospheric activity and/or very low coronal
  illumination because the line becomes so optically thin that the
  photospheric continuum shines through.

Telescopes observing in \ion{He}{i} are CHIP \citep{Elmore1998} at MLSO and the
OSPAN instrument \citep[formerly ISOON, see][]{Neidig1998} at the
National Solar Observatory at Sacramento Peak. The latter also provides
observations in H$\alpha$.

The Chromospheric Telescope \emph{ChroTel} introduced here records
data in all three spectral lines (\ion{Ca}{ii} K, H$\alpha$,
\ion{He}{i} 10830~\AA) with a cadence as fast as one minute. 
This is done almost simultaneously, and as the different lines sample
different temperature regimes (and therefore heights), this gives
a more holistic view of phenomena in the chromosphere and their
connection to higher layers. In addition, the data in \mbox{\ion{He}{i}} are
taken at seven wavelength positions in and around the \mbox{\ion{He}{i}}
triplet with a tunable Lyot-type filter. As we later demonstrate, the line-of-sight (LOS)
velocities across the full disk can be determined with these filtergrams,
which is essential for investigating the dynamics in the chromosphere. Since it is
located on Tenerife, i.e., in a different time-zone from both
the CHIP and OSPAN instruments, ChroTel will also increase the
available temporal coverage of full-disk data in \ion{He}{i} 10830~\AA. 

The outline of the paper is as follows. Section 2 gives a brief
overview of the scientific scope of ChroTel, followed by a description of the
instrument and data in Sects. 3 and 4. Section 5 illustrates our method
for extracting information about line-shifts from the intensity filtergrams
in \mbox{\ion{He}{i}}. Finally, we make some concluding remarks on the
data quality and the potential of the method for the velocity
determination.  

\section{Scientific scope}
ChroTel observes a wide range of solar phenomena of scientific
importance, some of which are presented here. This is of course
neither complete nor representative of what can be achieved with the data,
but gives an impression of the potential of the instrument. 

\subsection{Filaments, flares, and Moreton waves}
Filaments (and prominences) are naturally a subject of chromospheric full-disk
observations both for statistical reasons and because they often occupy
a significant fraction of the solar disk. The measurement of their
LOS velocities permits the examination of large-scale flows within
filaments. Along with information about the magnetic field
configuration, this can serve as an important ingredient of filament
models, especially for the onset of flares and coronal mass ejections (CMEs). 

Related to this are Moreton waves \citep{Moreton1960},
which are horizontal waves traditionally observed in H$\alpha$ that propagate
away from flare sites at speeds of up to 2000~km\,s$^{-1}$. Authors
have tried to establish connections bet\-ween EIT waves  \citep[the
name stems from observations of the waves with the Extreme ultraviolet
Imaging Telescope, see][]{Thompson1998} higher up in the atmosphere and Moreton
waves, primarily on the basis of the co-spatiality of the two phenomena. While some authors indeed
find co-spatial features \citep[e.g.,][]{Thompson2000,Warmuth2002},
others come to a different result \citep[e.g.,][]{Eto2002,Okamoto2004}
and question whether a connection exists. 

Investigations of observations in an intermediate layer seem
favorable and have indeed been done with data in \mbox{\ion{He}{i}}
10830~\AA. \cite{Vrsnak2002} find co-spatiality in H$\alpha$,
\mbox{\ion{He}{i}}, and EIT maps, with a forerunner in
\mbox{\ion{He}{i}} with respect to H$\alpha$. They conclude that the waves are of a mechanical nature,
whereas \cite{Gilbert2004} suggest that the co-spatial features they
find in \mbox{\ion{He}{i}} with respect to EIT data indicate a `chromospheric imprint' of
a wave in the corona triggered by the enhanced coronal EUV
emission, which then again leads to an increase in the \mbox{\ion{He}{i}}
absorption. With almost co-temporal data in both H$\alpha$ and
\mbox{\ion{He}{i}}, this controversy can be investigated using ChroTel
data. As \cite{Gilbert2004} conclude in their paper: ``In any case, a
high-time cadence \mbox{\ion{He}{i}} $\lambda$10830 data set should provide an
additional useful tool in the study of coronal waves."

The cadence of 3~minutes for the synoptic observations is,
  however, not perfectly suited to investigating short-lived events such as
  flares or waves travelling at speeds of up to 2000~km\,s$^{-1}$. This would have to be done in a
  dedicated mode of operation looking for these events, e.g.,
  by taking images with a high cadence only in the line cores of
  H$\alpha$ and \ion{He}{i}. Since Moreton waves travel a significant
  distance during the acquisition time of about 8~s for the whole \ion{He}{i}
  filtergram sequence, the determination of LOS velocities in
  \ion{He}{i} is of limited use for this purpose, but the
  propagation of the waves can also be traced in high-cadence
  intensity images, e.g., using the differences between consecutive
  images.

\subsection{Supersonic downflows}
Magnetic flux in the quiet Sun appears in two different shapes:
diffuse and weak magnetic flux in the internetwork, and concentrated
magnetic field in both the network and isolated flux tubes. A possible
method for the transition from the diffuse to the concentrated state
was proposed by \cite{Parker1978}, which he called the
\emph{super\-adiabatic effect}: when the magnetic field lines are shuffled 
to and `squeezed' at the boundaries of supergranules, the magnetic
field strength rises and suppresses convection. In these regions, this leads to a
cooling of the plasma which is already in a
general downdraft region. The downdraft is further enhanced by the
cooling, through which plasma in the upper parts of a filled flux tube
can lose its support from below and the flux tube becomes evacuated
rapidly. The plasma from above should then (nearly) undergo a
free-fall motion in this picture, which is usually referred to as
\emph{convective collapse}. These motions indeed were observed in
a plage region by \cite{Schmidt2000} in H$\beta$ and \mbox{\ion{He}{i}}
10830~\AA, with plasma movements at a constant acceleration of
200~m\,s$^{-2}$. The motions lasted for about 90~s, which agrees with the results of
\cite{Grossmann1998} who inferred a duration of several minutes for
convective collapses from 2D simulations. It is puzzling, however, that
the reports of these events are very sparse, although Parker's
scenario should be rather general and occur essentially on an everyday
basis for the Sun.  

\cite{Lagg2007} also report high-velocity downflows in
\mbox{\ion{He}{i}}, but in the vicinity of a growing pore. The
downflows were lasting (and increasing) for over an hour, so the
convective collapse scenario in this case is rather unlikely. The
authors interpreted their results as a continuous  drainage of a
slowly rising magnetic flux tube, leading to increasing downflows in
the footpoints of the loop. 

With ChroTel, it will be possible to improve the quality of the statistics of the
rates and properties of these events. The downflows in the aforementioned
observations were seen in regions extending from about 5 to 30
arcseconds, with velocities up to
42~km\,s$^{-1}$. In Sect.~\ref{sec:doppler}, we demonstrate that this should
be easily accessible with ChroTel data in \mbox{\ion{He}{i}}.     

\subsection{Onset of the fast solar wind in the chromosphere?}
It has been known for many decades that the solar wind exhibits a slow and a
fast component, the latter having average speeds of 700-800~km\,s$^{-1}$
\citep[e.g.][]{Phillips1995}. It is also known that the fast component
emerges from coronal holes \citep[e.g.,][]{Krieger1973}, i.e., regions
appearing darker than their surroundings at EUV and X-ray wavelengths,
although there are indications that it might also emerge from quiet
Sun regions \citep{Habbal1997}. 

It is still unclear, however, at which height and temperature the
acceleration of the fast solar wind sets in. \cite{Dupree1996}
investigated wing asymmetries in \mbox{\ion{He}{i}} 10830~\AA$\;$profiles,
i.e., in the upper chromosphere and lower transition region, from
scans covering two polar coronal holes. The profiles in the coronal holes
were found to be blueshifted, with a center-to-limb variation
indicating a line-of-sight effect for the velocities, i.e., the
amplitudes of the velocities were found to be dependent on the cosine
of the heliocentric angle. This was interpreted as a radial outflow in
the coronal holes and as the onset of the fast solar wind at
chromospheric heights.     

\begin{figure*}
  \centering
  \includegraphics[width=\textwidth]{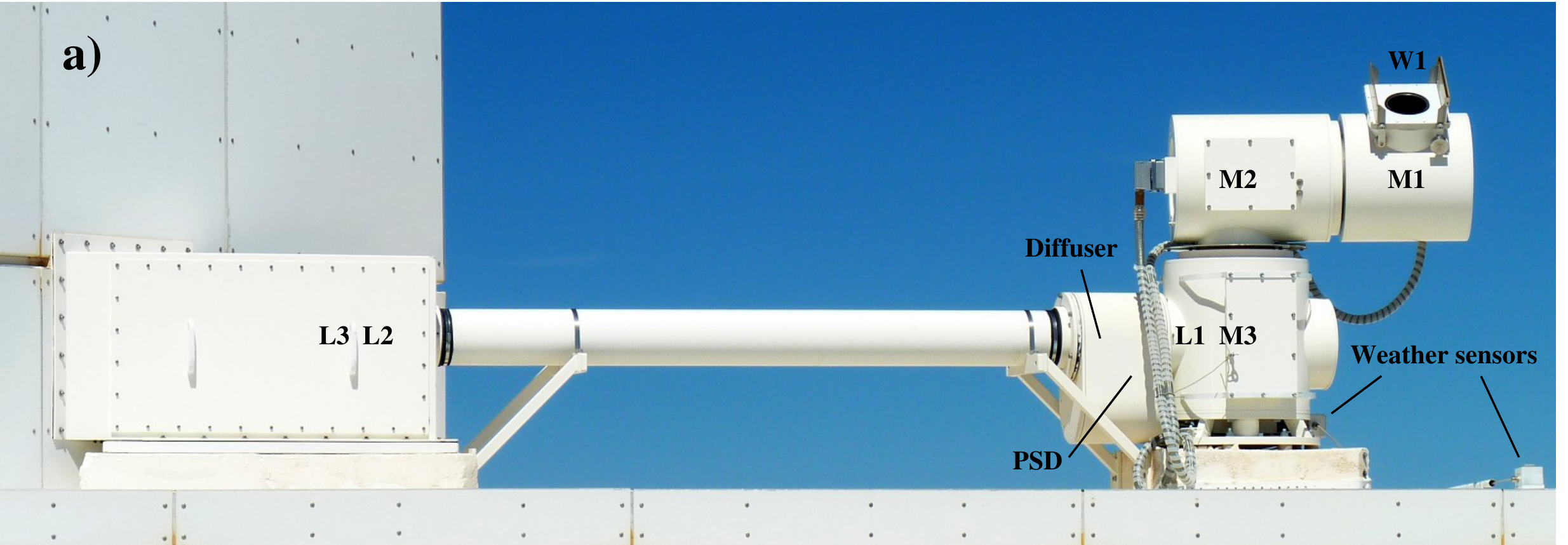}\\[10pt]
  \includegraphics[width=\textwidth]{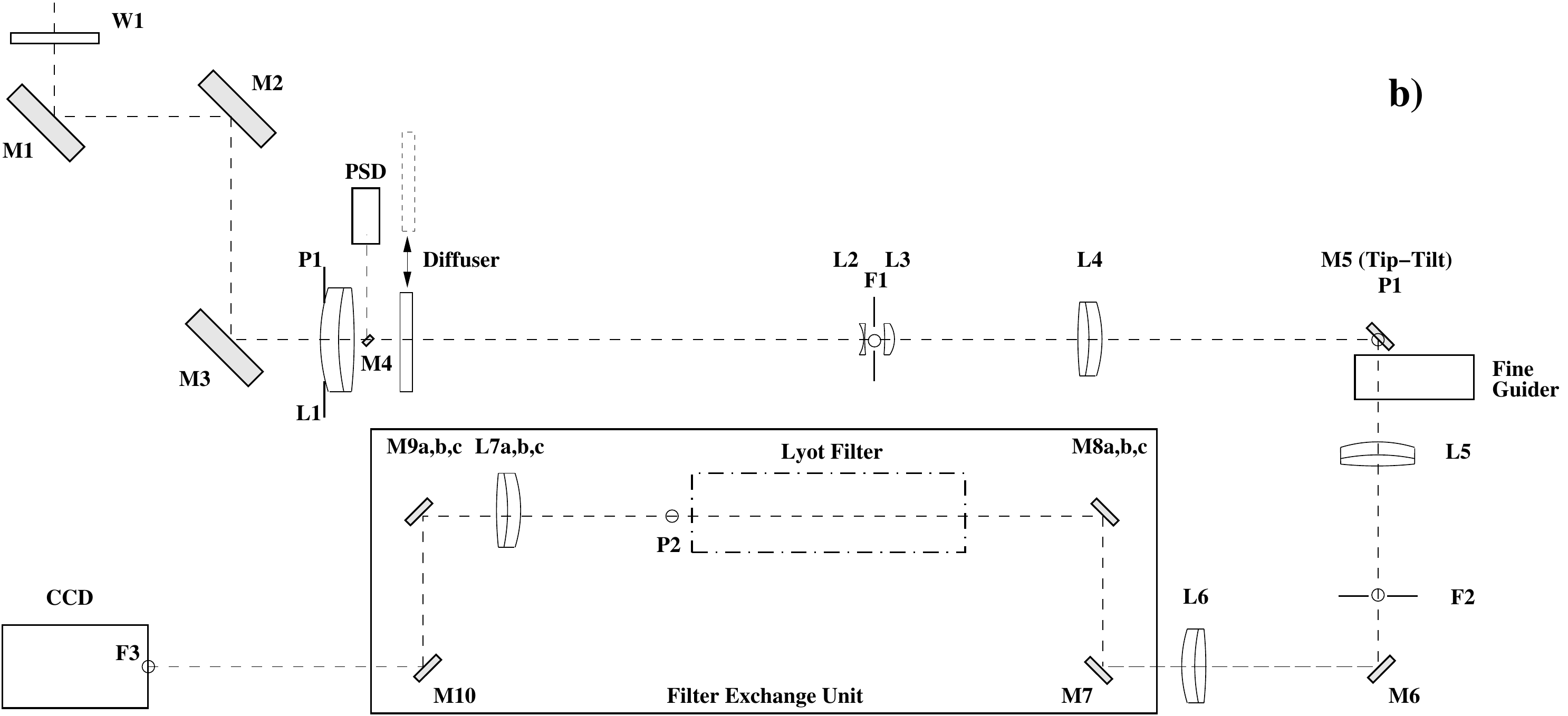}
  \caption{a) Telescope structure outside the optical lab. b)
    Schematic overview of the optical path (not to scale): W1
    entrance window, M1-3 telescope mirrors, L1 telescope lens with
    aperture, PSD position sensing detector, L2-3 field lenses,
    L4-7a,b,c achromatic lenses, M5 tip-tilt mirror, M7/M10 rotatable
    mirrors, M8-9a,b,c static mirrors, F1-3 focal planes, and P1-2
    pupil planes.} 
  \label{fig:telescope}
\end{figure*}

\cite{Tu2005} investigated SUMER \citep{Wilhelm1995} scans of a polar
coronal hole in the emission lines of \ion{Si}{ii}, \ion{C}{iv}, and 
\ion{Ne}{viii}. The \ion{Ne}{viii} emission forming in the lower corona
at 6~$\times$~10$^5$~K was found to be blueshifted on average by
9~km\,s$^{-1}$, whereas in the \ion{C}{iv} emission that is formed at
1~$\times$~10$^5$~K, only weak signatures of outflows were seen at
the disk center \citep{Dere1989} and at the poles
\citep{Peter1999a,Peter1999b}. The question of course arises of how
the blueshifts in \mbox{\ion{He}{i}} 10830\,\AA$\;$can depict the onset of the fast
solar wind when only weak outflows are seen in \ion{C}{iv} higher up in the
atmosphere.  

Along with co-temporal velocity information about the transition region,
ChroTel can help us to determine whether the outflows seen by
\cite{Dupree1996} are indeed a signature of the fast solar wind in the
chromosphere or should be interpreted in a different
way. As we show in Sect.~\ref{sec:doppler}, coronal holes pose a challenge to
the velocity determination with ChroTel filtergrams because the
intrinsic \ion{He}{i} absorption is very weak in these
regions. However, this might be addressed with the averaging of
data because coronal holes are long-lived pheno\-mena.

\section{The instrument}
The configuration of the instrument is described in detail in
\cite{Kentischer2008} and \cite{Halbgewachs2008}, so only a brief
overview is given here. 

\subsection{Technical design and guiding}

The outdoor structure of the telescope is a domeless turret in an
alt-azimuth mount. The turret is sealed to protect the optical
elements from wind, humidity, and extreme
temperatures. Figure~~\ref{fig:telescope}\,a) shows the telescope
structure attached to the building of the Vacuum Tower Telescope (VTT)
on Tenerife.

Because of the alt-azimuth mount both the azimuth and the ele\-vation
axis have to be rotated in order to follow the position of the Sun
during the day. The guiding is conducted in three consecutive steps.
First, the values for azimuth and elevation of the Sun on the sky are
calculated from ephemerides, not taking into account atmospheric
refraction. Second, a position-sensing detector (PSD) corrects small
and slowly varying deviations of the Sun's designated position within
a field-of-view (FOV) of 2.2$^{\circ}$. The last step is the compensation of fast image
movements with a tip-tilt mirror (M5 in
Fig.~\ref{fig:telescope}\,b)). It operates with a closed control loop
at a cycle time of 108\,$\muup$s and within a FOV of
0.7$^{\circ}$. The pointing accuracy of this system is better than
0.5\,arcsec.  
\begin{table*}
\center
\caption{Overview of the observed spectral lines. The values in italics indicate the
  default settings for the synoptic observations.\label{table:channels}}  
\begin{tabular}{lccccc}
  \hline\hline
  Spectral line & wavelength & Lyot filter widths & typical exposure & \multicolumn{2}{c}{minimum cadence [s]}\\
  & [\AA] & FWHM [\AA] & time [ms] & single channel & all channels\\[2pt]
  \hline\\[-7pt]
  \ion{Ca}{ii} K & 3933.7 & {\it 0.3}; 0.6; 1.2 & 1000 & 10 & 60\\[2pt]
  H$\alpha$ & 6562.8 & {\it 0.5}; 1.0 & 100 & 10 & 60\\[2pt]
  \mbox{\ion{He}{i}} & 10830.3 & {\it 1.3} & 1000 & 30 & 60\\[2pt]\hline
\end{tabular}
\end{table*}
\subsection{Optics}
The light enters through a wedged and anti-reflection-coated entrance
window (W1). The window is protected by a shutter when the instrument
is not operating to minimize contaminations by dust and humidity. The
light is redirected to the optics lab inside the VTT
building by the flat telescope mirrors \mbox{M1-3} and imaged onto a single CCD by
a combination of achromatic lenses (L1: f~=~2250\,mm, 100\,mm aperture;
L4: f~=~800\,mm; L5: f~=~600\,mm; L6: f~=~500\,mm; L7: f~=~800\,mm) and
field lenses (L2-3). The aperture of 100\,mm leads to a theoretical
resolution limit of 1\,arcsec for the calcium channel, 1.7\,arcsec
for H$\alpha$, and 2.7\,arcsec for \mbox{\ion{He}{i}}. The spatial
sampling on the 2048x2048~pixel CCD is
about 1\,arcsec per pixel, i.e., in
the first two channels, the spatial resolution is sampling limited to
about 2\,arcsec by the detector. 

All channels are recorded with one camera, hence the CCD must be
sensitive over a large wavelength range from 393\,nm to 1083\,nm. A
Kodak KAF-4320E CCD was selected, operated in a water-cooled 
Spectral Instruments Series 800 camera. The detector has a dynamic range of
14\,bit, a pixel size of 24x24\,$\muup$m, and a quantum efficiency (QE) of
around 35\% for \mbox{\ion{Ca}{ii} K}, 63\% for H$\alpha$, and 2\% for \mbox{\ion{He}{i}}. 

To perform observations in all three spectral lines, the beam has to be
redirected successively to three Lyot-type filters (see Section \ref{sec:lyot}). For this,
an in-house developed \emph{filter exchange unit (FEU)} is used: the filters
are mounted in an equilateral triangle setup with static flat mirrors
both at the entrance and the exit of the filters (M8-9a,b,c
in Fig.~\ref{fig:telescope}\,b)). These mirrors are tilted by 45$^{\circ}$
with respect to the optical axis of the filters. Two rotatable mirrors
(M7 and M10), also tilted by 45$^{\circ}$, redirect the light to one
of the static mirrors sets, hence through one of the Lyot
filters. Rotating the (coupled) mirrors M7 and M10 by 120$^{\circ}$
around the axis coinciding with the direction of the incident light
then allows a different filter to be selected. The mirrors M7-10 have a reflectivity
of 90\% at the relevant wavelengths, i.e., the FEU entails a loss of
more than one-third of the incident intensity. Along with the low QE of the
detector in the \ion{Ca}{ii} K and \mbox{\ion{He}{i}} wavelengths, this involves comparably
long exposure times of up to 1~s in these channels as shown in Table~\ref{table:channels}. 

Flatfielding is done using a diffuser plate that can be
inserted into the beam when necessary. The diffuser plate is
custom-made and generates a uniform intensity over a larger angular
range than regular holographic diffuser plates. It is mounted within
the outside telescope structure to take care of contaminations on
all subsequent imaging optics. The diffuser plate rotates during
the flatfield exposures to compensate for possible inhomogeneities in
the manufacturing process. To do this, the period of the
rotation is chosen as an integer multiple of the exposure
time. Flatfields are currently taken once per hour in every channel.

\subsection{Lyot filters\label{sec:lyot}}
For each of the spectral lines, a Lyot-type birefringent narrow-band
filter is used for the observations. All filters contain wide-fielded
elements for a larger monochromatic acceptance angle. We now provide a
short description of each Lyot filter:\\[-8pt]

\begin{figure}
  \centering
  \resizebox{0.955\hsize}{!}{\includegraphics[width=\textwidth]{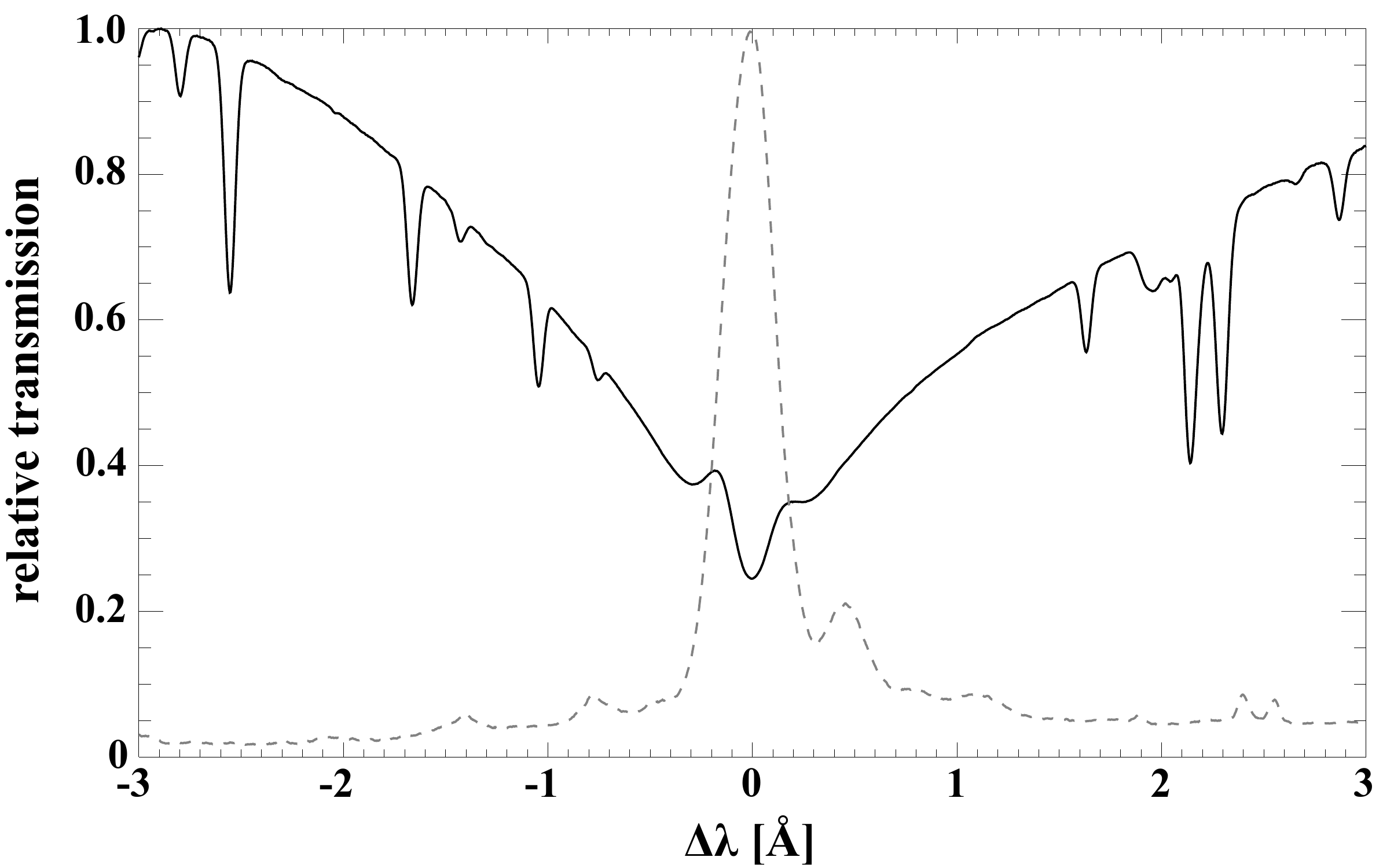}}
  \caption{Measured passband (grey dashed line) of the Lyot filter
  used for ChroTel \mbox{\ion{Ca}{ii} K} observations. The zero point marks the
  minimum of the \mbox{\ion{Ca}{ii} K} line at 3933.7~\AA. For comparison, a
  spectrum around the \mbox{\ion{Ca}{ii} K} line is overplotted (solid line,
  recorded with the spectrograph of the VTT).}  
  \label{fig:ca_transm}
\end{figure}

\ion{Ca}{ii} K:\, The wavelength of the filter passband is adjusted to
the central minimum of the \ion{Ca}{ii} K line at 3933.7\,\AA. The FWHM of the
filter passband in the narrowest possible mode is 0.3\,\AA, which is
the default setting for the synoptic observations (see Table
\ref{table:channels}). This is broad enough to observe the emission of
both the K$_{\textrm{\small 2\normalsize r}}$ and K$_{\textrm{\small
    2\normalsize v}}$ emission peak up to about 0.25\,\AA\,on the red
and blue side of core of the line. Detachable polarizers at the entrance and
the exit of the filter also allow filter passbands with a FWHM of
0.6\,\AA~and 1.2\,\AA. Figure~\ref{fig:ca_transm} shows a
  measurement from 2011 of the narrowest possible filter passband. The
  FWHM of the central transmission peak determined from a Gaussian fit to the curve is 
  0.29~\AA. The measurement also revealed a slightly enhanced
  contribution from a side lobe in the red wing of the
  \mbox{\ion{Ca}{ii} K} line.\\[-8pt]   

\begin{figure}
  \centering
  \resizebox{0.95\hsize}{!}{\includegraphics[width=\textwidth]{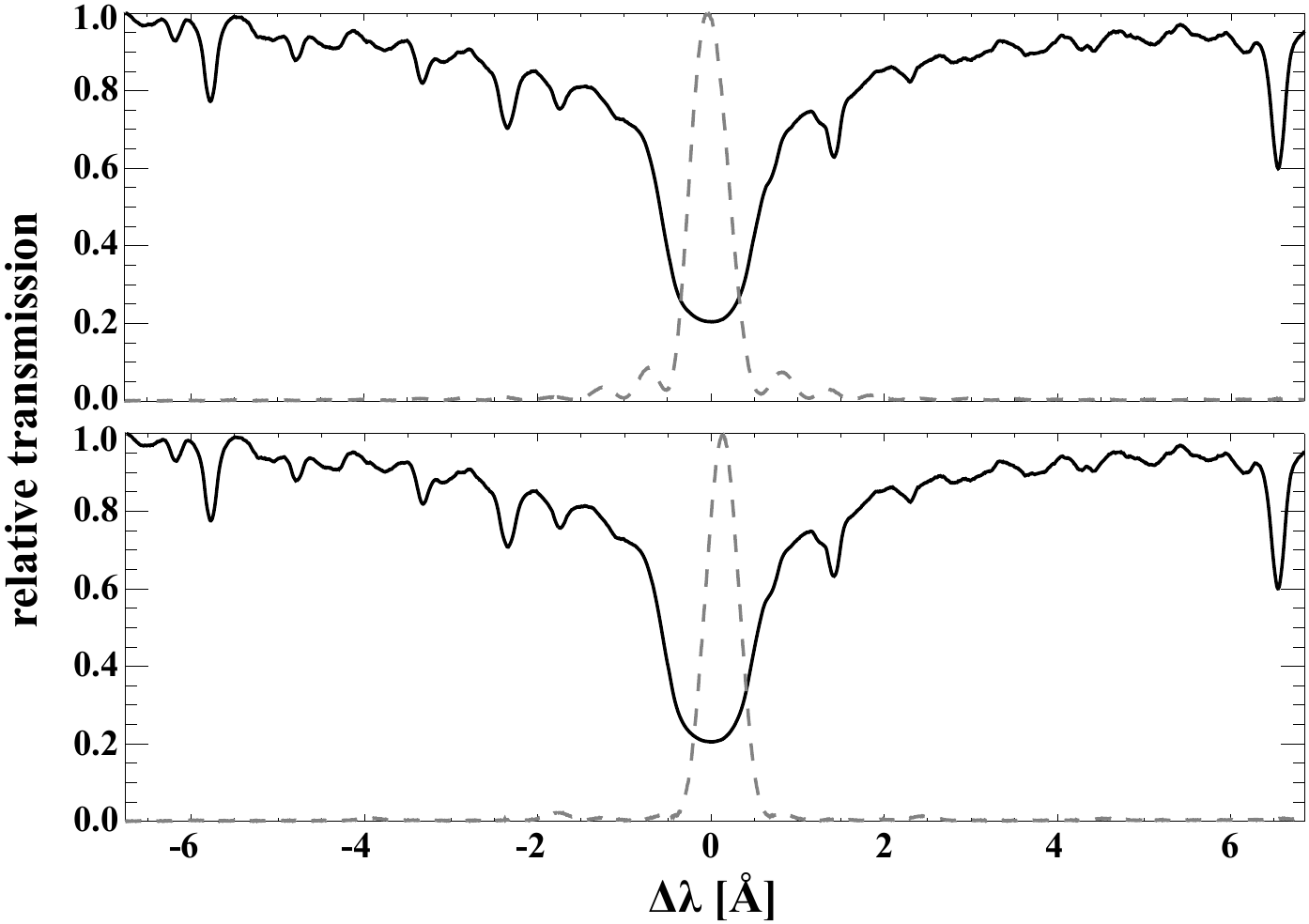}}
  \caption{Measured passbands (grey dashed lines) of the Lyot filter
  used for ChroTel H$\alpha$ observations. The zero point marks the
  minimum of the H$\alpha$ line at 6562.8~\AA. \emph{Top:} filter passband
  without contrast element. \emph{Bottom:} filter passband with contrast
  element. For comparison, a spectrum around the H$\alpha$ line is
  overplotted (solid lines, recorded with the spectrograph of the
  Schauinsland Observatory).}
  \label{fig:ha_transm}
\end{figure}

H$\alpha$:\, The FWHM of the filter passband in the default
setting is 0.5\,\AA. It can be widened to 1.0\,\AA~when a detachable entrance
polarizer is removed. An additional contrast element can be used to
suppress side lobes of the transmission
curve. Figure~\ref{fig:ha_transm} shows a measurement of the
transmission curve 
from 2009 without and with the contrast element. In the latter case, the side lobes
are significantly reduced, whereas the central wavelength of
the transmission peak is shifted slightly towards the blue by
0.1\,\AA. This is nevertheless the preferable (and default) setting
because the wings of the line are so bright compared to the core that
the fraction of the intensity coming from the side lobes is over 70\%
when the contrast element is not used. With the contrast element, this
drops to 40\%, i.e., despite the shift, the largest fraction
of the signal comes from the core of the line. The FWHM of the central
transmission peak determined from a Gaussian fit is 0.48~\AA.\\[-8pt]   

\begin{figure}
  \centering
  \resizebox{0.99\hsize}{!}{\includegraphics[width=\textwidth]{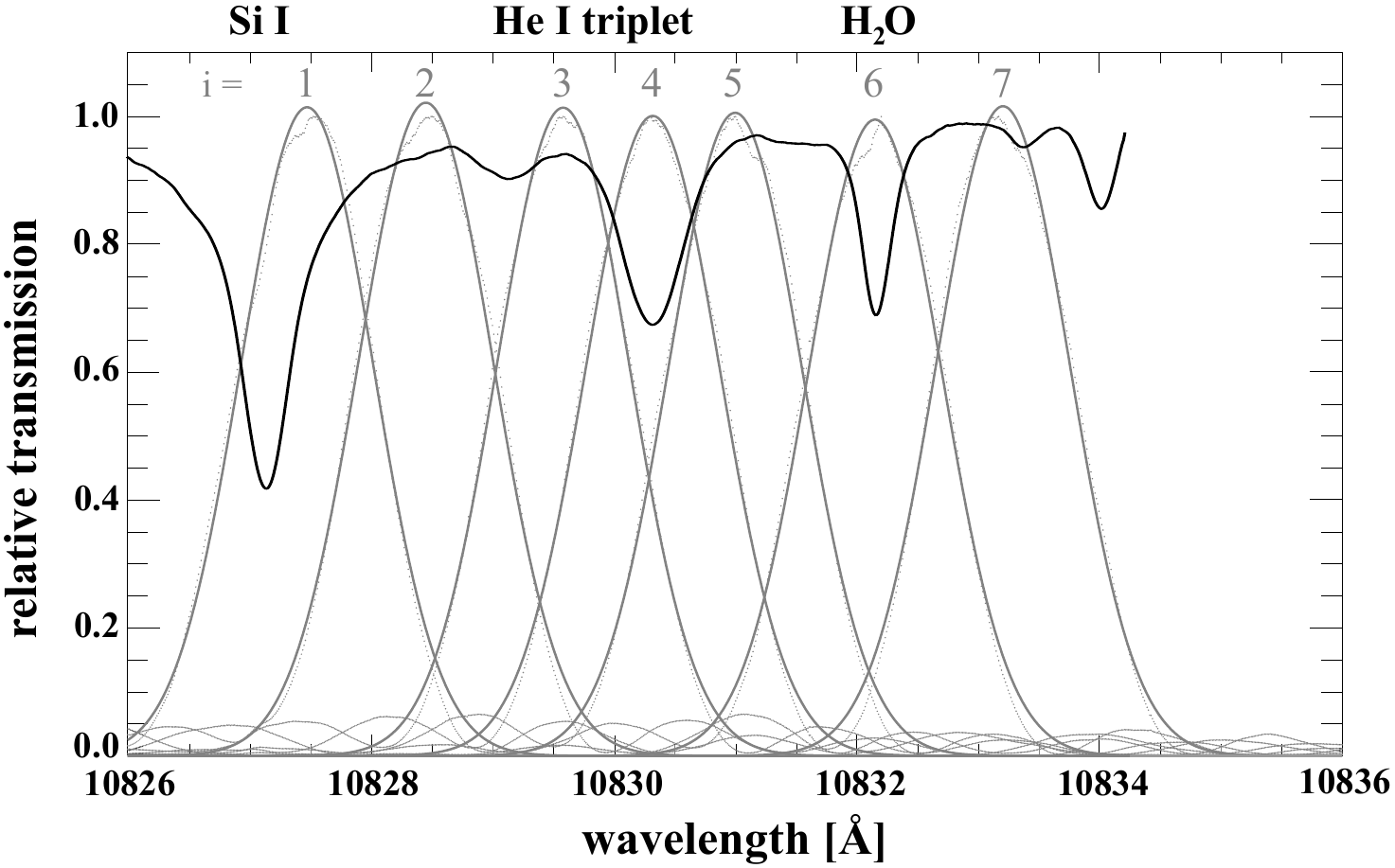}}
  \caption{Measured passbands of the tunable Lyot filter used
    for ChroTel \mbox{\ion{He}{i}} 10830\,\AA\,observations. \emph{Grey dotted
      lines:} measured values. \emph{Grey solid lines:} polynomial fit
    to the measurements. Central wavelength positions for the
    passbands in \AA ngstroms (\AA), from left to right: 10827.45, 10828.47,
    10829.60, 10830.30, 10831.00, 10832.13, 10833.15. For comparison,
    a spectrum around the \mbox{\ion{He}{i}}
    triplet from a plage region is overplotted in
    black (smoothed; observed with the Tenerife Infrared Polarimeter).} 
  \label{fig:he_pb}
\end{figure}

\mbox{\ion{He}{i}} 10830\,\AA:\, A tunable filter is used for the
observations in the \mbox{\ion{He}{i}} infrared triplet. It was assembled by the
High Altitude Observatory based on the design of the filter for the CHIP instrument
\citep{Elmore1998}. The filter contains liquid crystal variable retarders
(LCVR) between the birefringent stages, which allow a rapid adjustment
of the central wavelength position of the transmission peak. CHIP
acquires filtergrams in and around the \mbox{\ion{He}{i}} line at seven fixed wavelength positions in a 
non-equidistant order. These wavelengths were also chosen for ChroTel to ensure
the comparability of the data of the two
instruments. In Sect.~\ref{sec:doppler}, we show that these filtergrams
can be used to determine line-shifts in \ion{He}{i} 10830~\AA,
hence chromospheric LOS velocities. Figure~\ref{fig:he_pb} shows a
measurement of the seven transmission curves of the \ion{He}{i} Lyot
filter from 2006. The overplotted \ion{He}{i} profile comes
  from a plage region. With its enhanced line depth, it shows a favorable case for the determination of
velocities. In the quiet Sun, the line depth is substantially smaller
than for the profile shown here.

The mean
FWHM of the passbands obtained from polynomial fits to the data is 1.29\,\AA. Shifting
the central wavelength position with the LCVRs is done in less than
60\,ms, i.e., with a typical exposure time of 1000\,ms, all
filtergrams are obtained within less than 8\,s. 

\section{Operation}
ChroTel was designed as a monitoring instrument for synoptic
observations. It operates autonomously and continuously, weather permitting, and can
also be remote-controlled. The instrument starts operating
automatically in the morning and shuts down in the evening, or in case
of unsafe conditions, which are being recognized based on data
from weather sensors for wind, humidity, and brightness. As soon as
the conditions are safe again after the instrument has shut down,
observations continue automatically.

All involved hardware components are controlled by a telescope
control software written in LabView. The software allows one to initiate
actions such as taking flatfields, choosing channels, and defining exposure
times etc. within a text script. By default, a standard script is run
for the regular operation described in Sect.~\ref{sec:synoptic}. Other
scripts can be used to customize the mode of operation, e.g., to
coordinate observations with the nearby spectrograph at the
VTT. Requests for specific modes of operation can be addressed to the
Kiepenheuer Institute\footnote[1]{Send observation requests to: \,
  chrotel-inquiry$@$kis.uni-freiburg.de} (Freiburg, Germany).
While no limitations on observation requests are in effect at
  the moment, requests for high-cadence observations over extended periods of
time might be denied or limited in order not to endanger the usefulness of
the synoptic program.

\subsection{Synoptic observations\label{sec:synoptic}}
An image is acquired in each channel once every three minutes in the
regular mode of operation. Small variations in the time consumed by
computer communications limit the precision of the cadence within one
channel to $\pm$1~s. Data in the three spectral lines cannot be recorded
simultaneously because the single camera allows only for one channel at a time. This
leads to a temporal offset of about 10-15~s between the channels.   

Single-channel observations can be done upon request with a minimum
cadence of 10\,s in \mbox{\ion{Ca}{ii} K} and H$\alpha$ and 30\,s in \mbox{\ion{He}{i}}. The minimum
cadence for observations in all three channels is one minute.

\subsection{Data reduction and availability}
All data reduction, processing, and transfer is automatized. Every acquired image
is instantly corrected for bias and gain with the most recent dark and
flatfield exposures. In practice, this means that first the dark current is subtracted from both
the flatfield and the observational data. All zero values
in the flatfield are then subsequently replaced by unity to avoid a division by
zero. A gain table is computed then from the flatfield, i.e., the pixel
values are rescaled with a division by the median value to a
range around unity to maintain the dynamic range of the
observational data. As a last step, the observational data are divided by the gain
table. The reduced data are then saved as 16-bit integer files in
gzipped FITS format. As a byproduct, live images are created as
1024x1024\,pixel JPG files, which are available a few minutes after
recording on the website of the Kiepenheuer Institute. They are also
used as a guiding aid for the nearby VTT and Gregor telescopes and
later on as preview images for the data. Observations acquired 300 days per
year and 7 hours per day would generate a yearly amount of data of
about 5\,TB.

Overnight, the data are transferred from Tenerife to Freiburg to make
them accessible for public use the next day. This includes the automated
creation of daily overview movies and webpages for all channels. Table
\ref{table:datprod} gives an overview of the provided data products
and the times they are available\footnote[2]{All data are open for scientific
usage and can be accessed at the following addresses:\\[4pt]
Live/most recent images:\\$\;\;$http://www.kis.uni-freiburg.de/index.php?id=457\&L=1\\[4pt]
Anonymous ftp access:\\$\;\;$ftp://archive.kis.uni-freiburg.de/pub/chrotel/\\[4pt]
Overview of available data:\\$\;\;$http://www.kis.uni-freiburg.de/${}_{\textrm{\symbol{126}}}$chrotel/index.html}.

\begin{table}
\center
\caption{Overview of the data products from ChroTel.\label{table:datprod}}  
\begin{tabular}{lcccl}
  \hline\hline
  Data type & Cadence & Resolution & Format & Availability \\
  & [min] & [pixels] &  & \\[2pt]
  \hline\\[-7pt]
  Scientific & 3 & 2048x2048 & FITS & Next day\\[2pt]
  Preview and & 3 & 1024x1024 & JPG &
  Few minutes \\
  live pictures & & & & after recording \\[2pt]
  Movies & 3 & 512x512 & AVI & Next day \\[2pt]\hline
\end{tabular}
\end{table}

\subsection{Instrument performance and data quality}
The robotic operation and automatic data processing has proven to
be reliable in the first one and a half years of
operation. Any outages were caused mainly by bad weather, maintenance, or
overall facility shutdowns in the winter.
\begin{figure*}
  \includegraphics[width=0.3333333\textwidth]{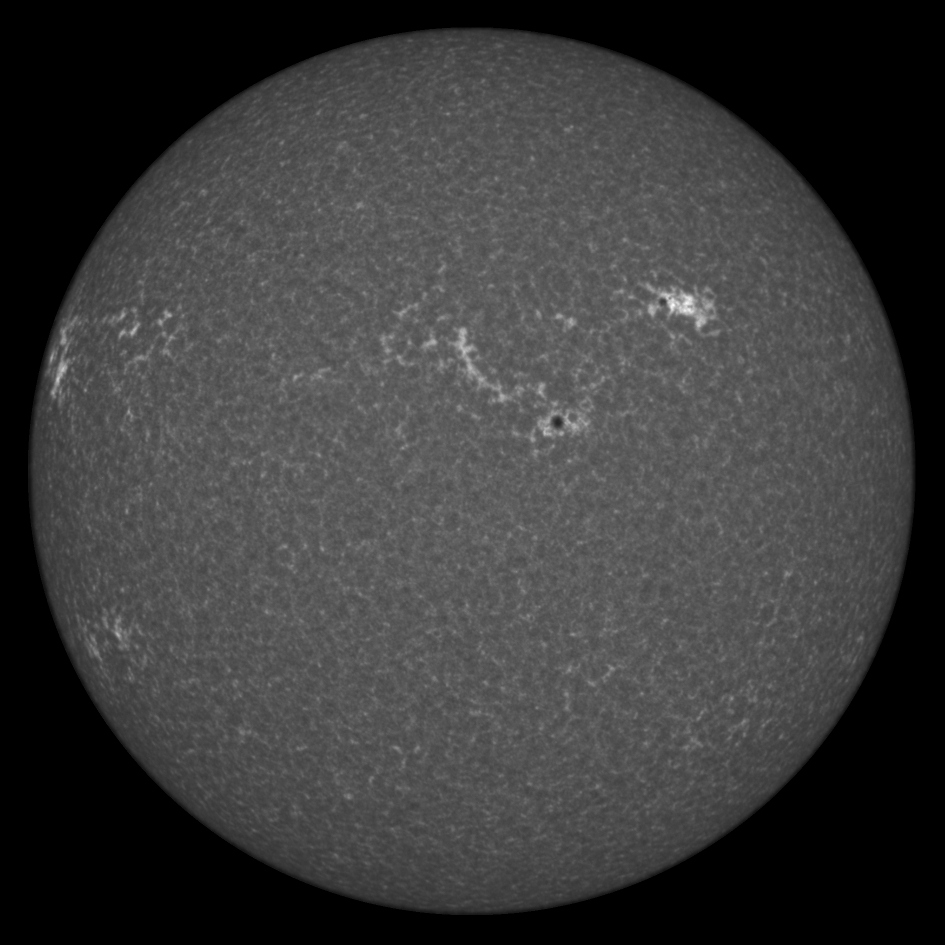}\includegraphics[width=0.3333333\textwidth]{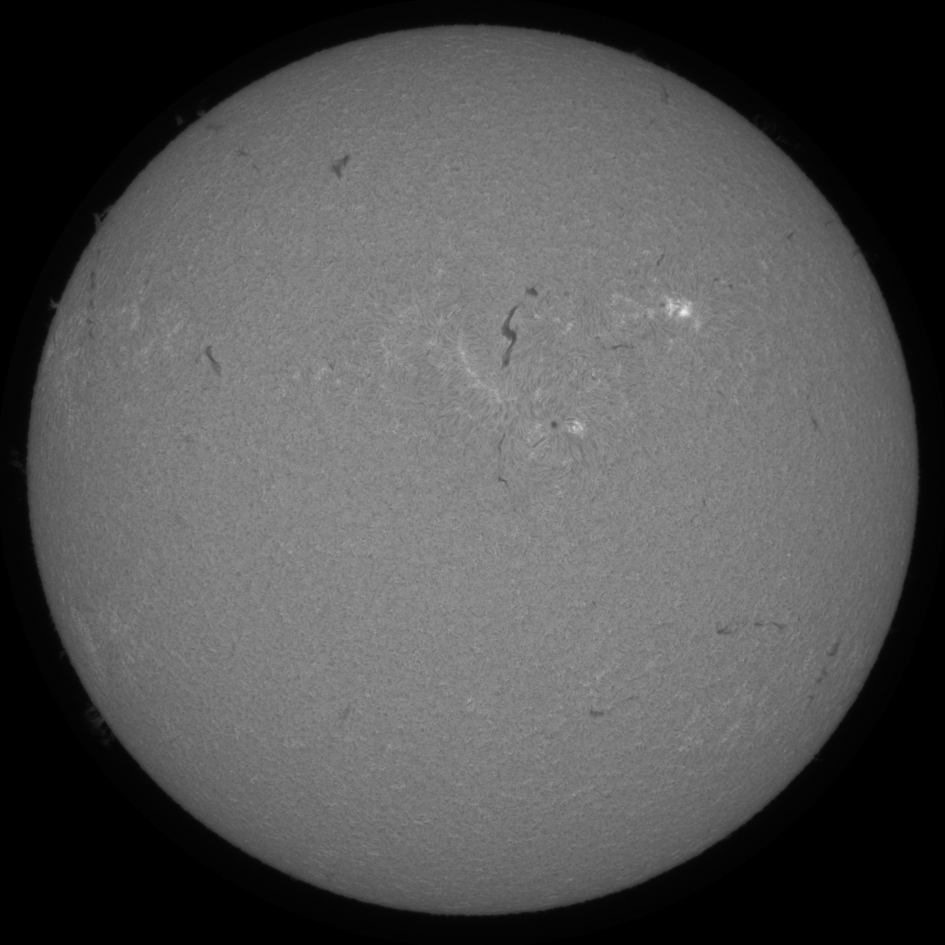}\includegraphics[width=0.3333333\textwidth]{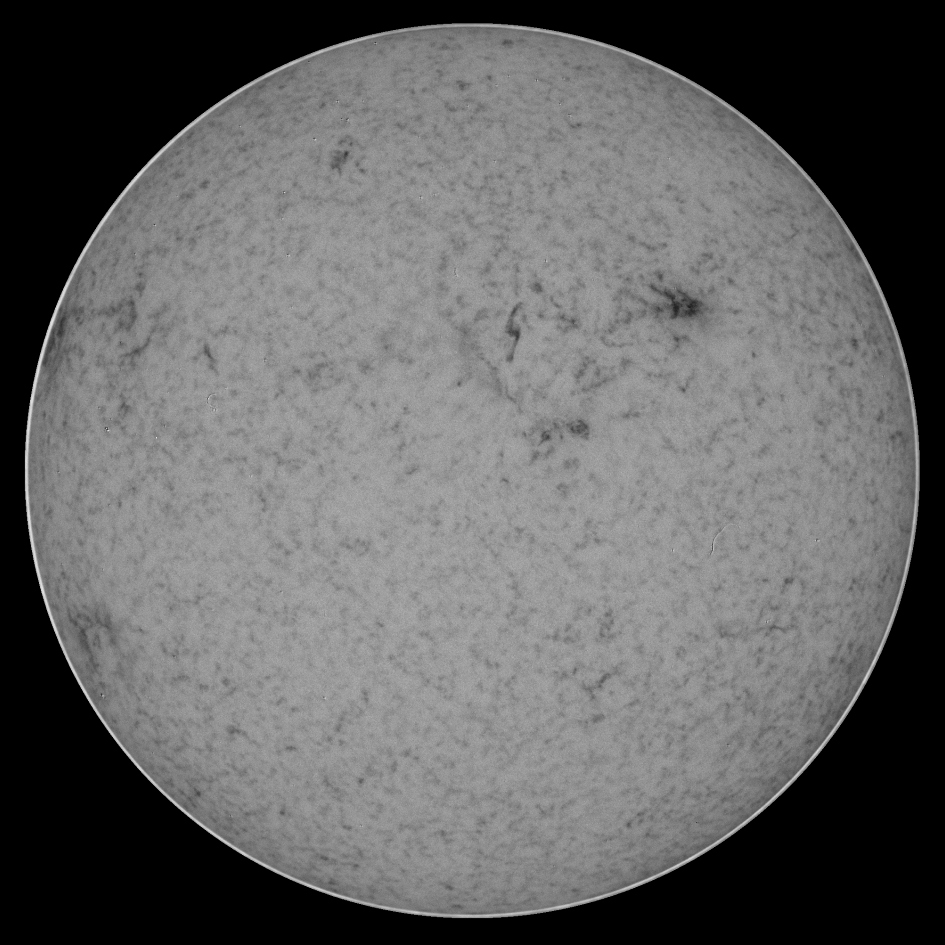}\\[0.3cm]\includegraphics[width=0.325\textwidth]{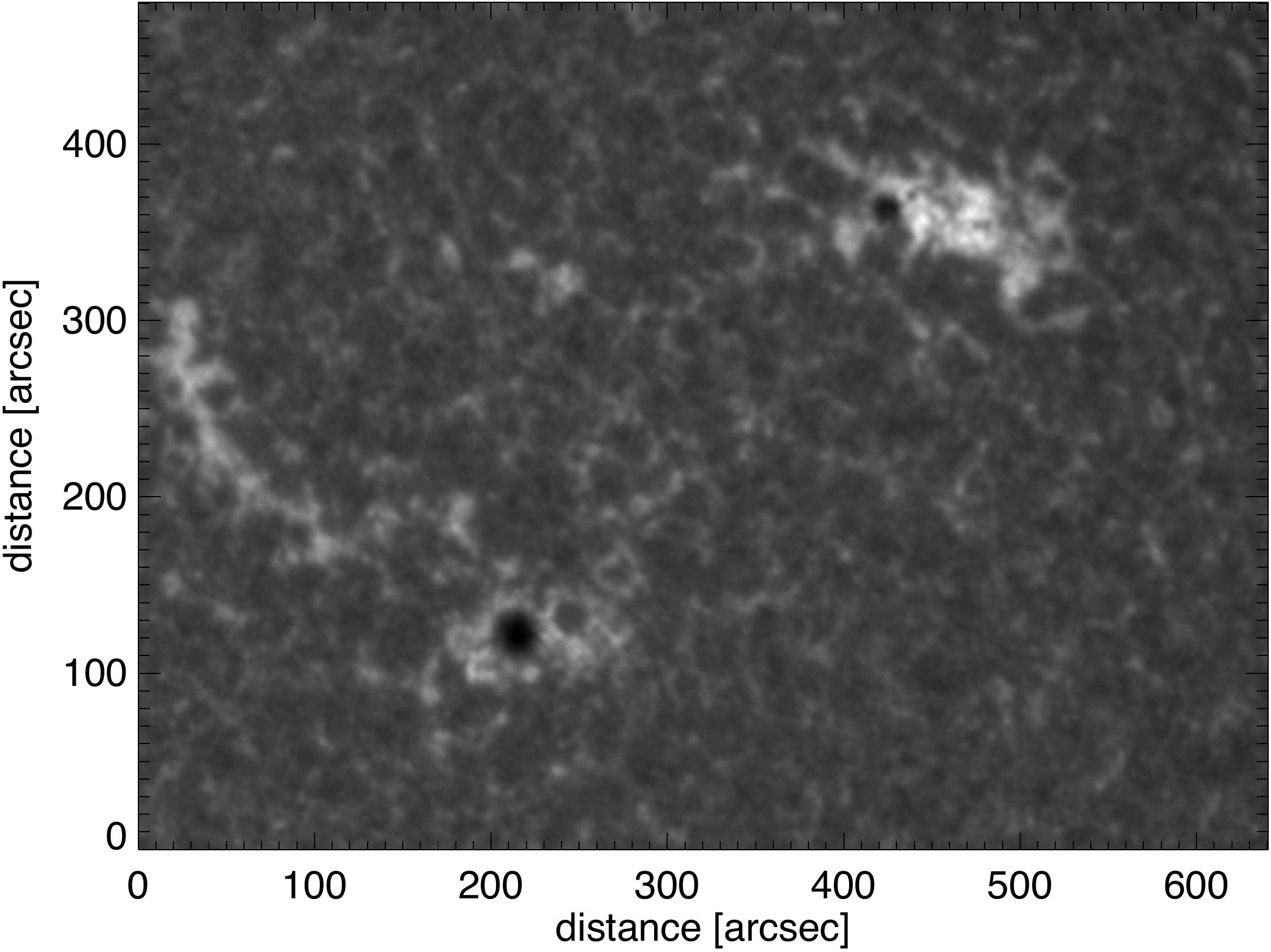}$\;\;\,$\includegraphics[width=0.325\textwidth]{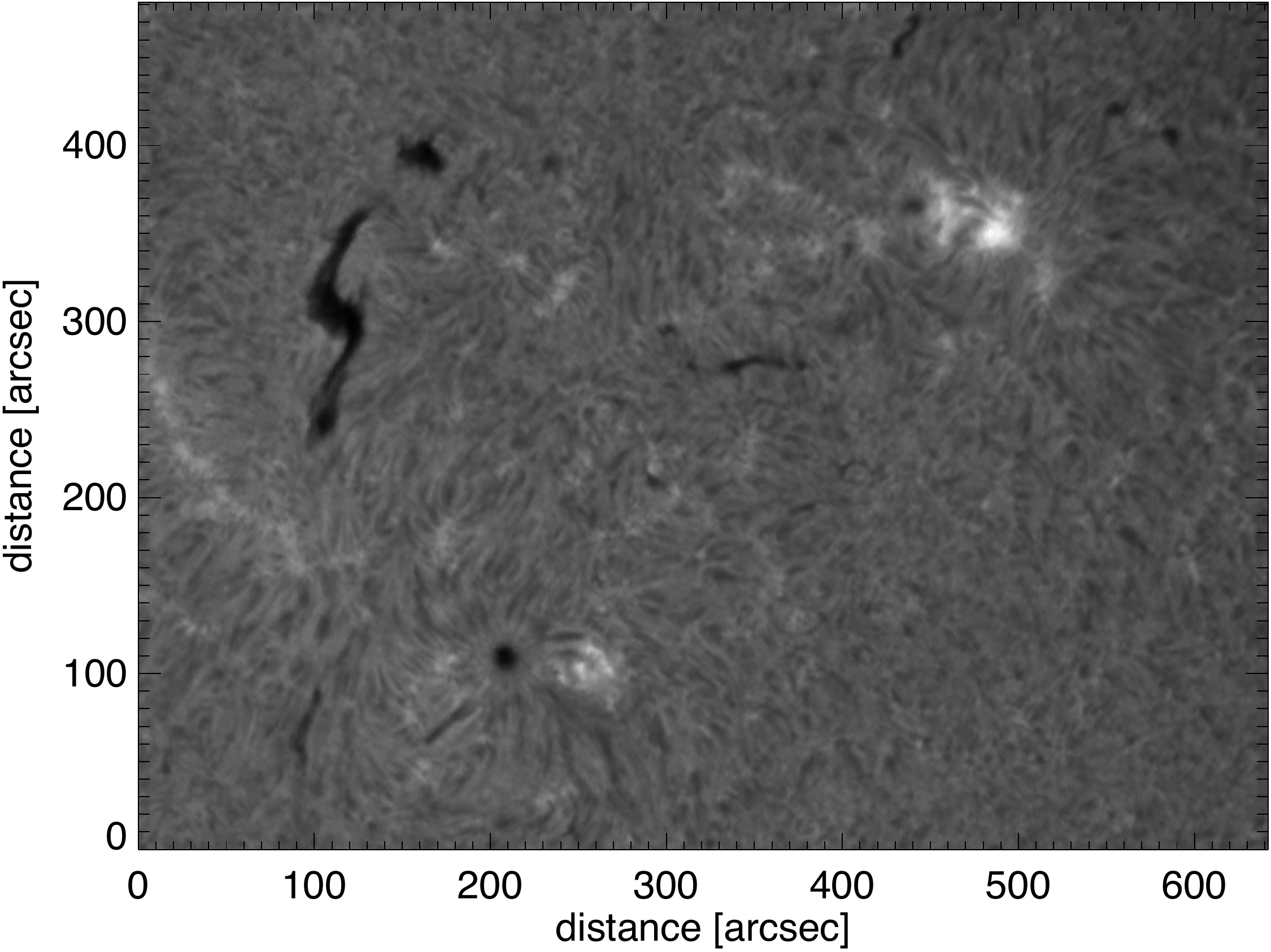}$\;\;\,$\includegraphics[width=0.325\textwidth]{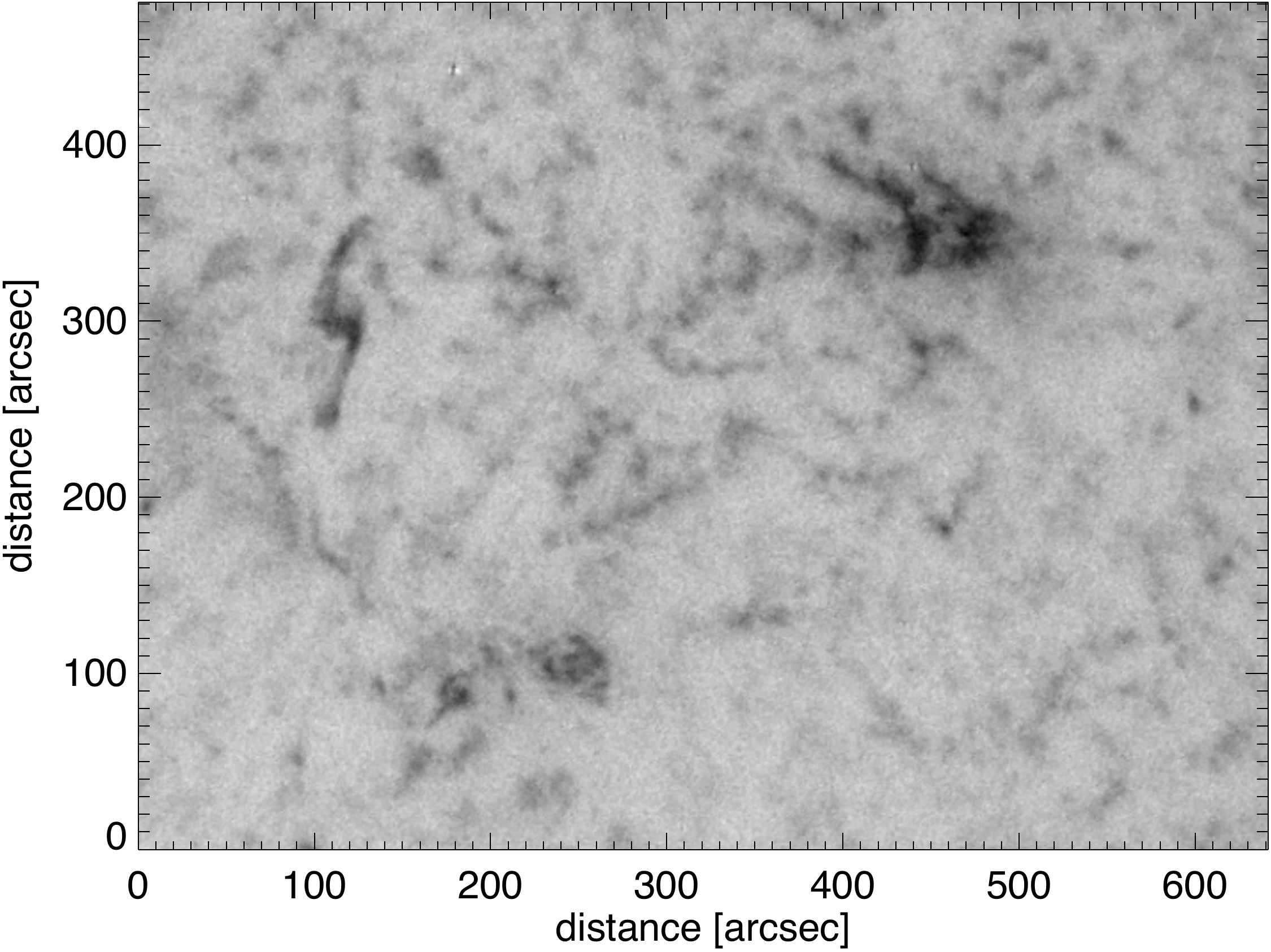}\\[0.2cm]\includegraphics[height=0.305cm]{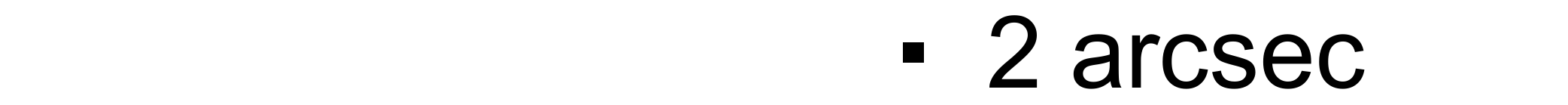}$\qquad\qquad\quad\,$\includegraphics[height=0.305cm]{two_arcsec}$\qquad\qquad\quad\,$\includegraphics[height=0.305cm]{two_arcsec}\\[-0.05cm]\includegraphics[height=0.285\textwidth]{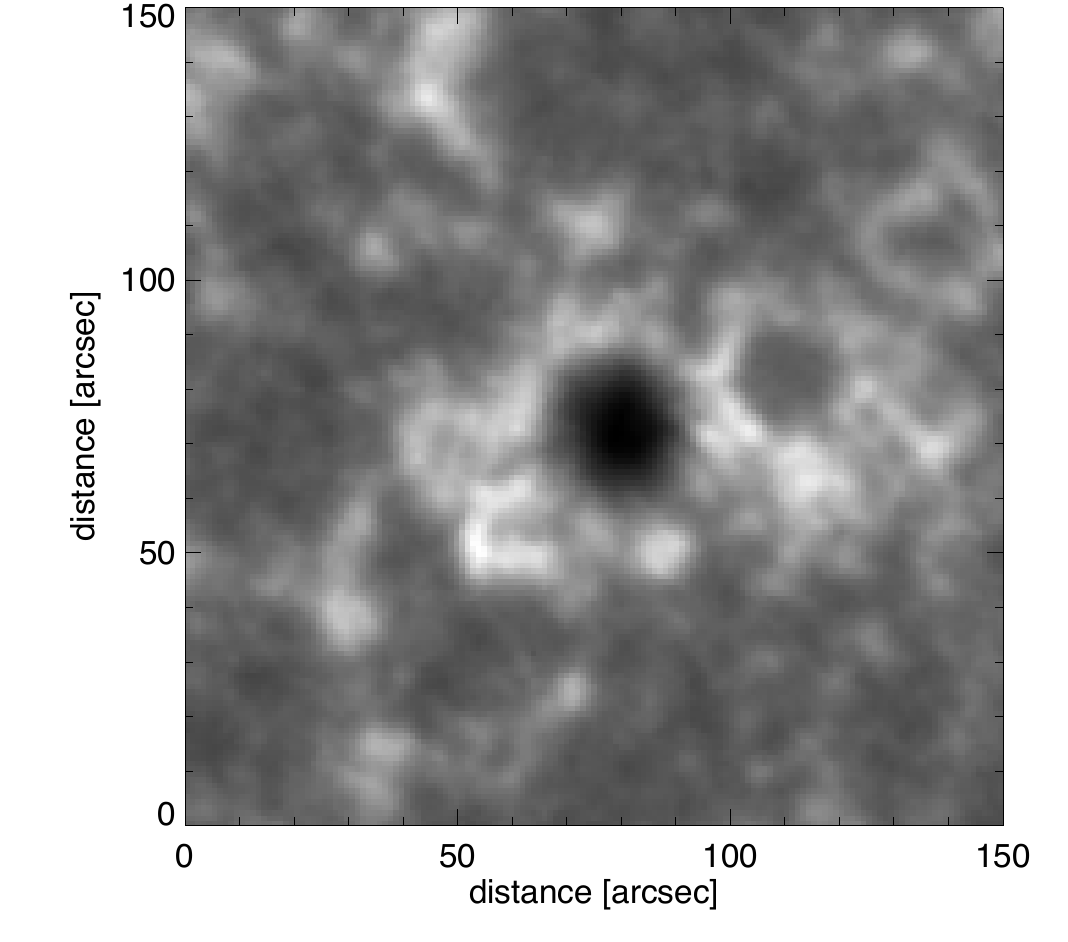}$\quad\;\,\,$\includegraphics[height=0.2851\textwidth]{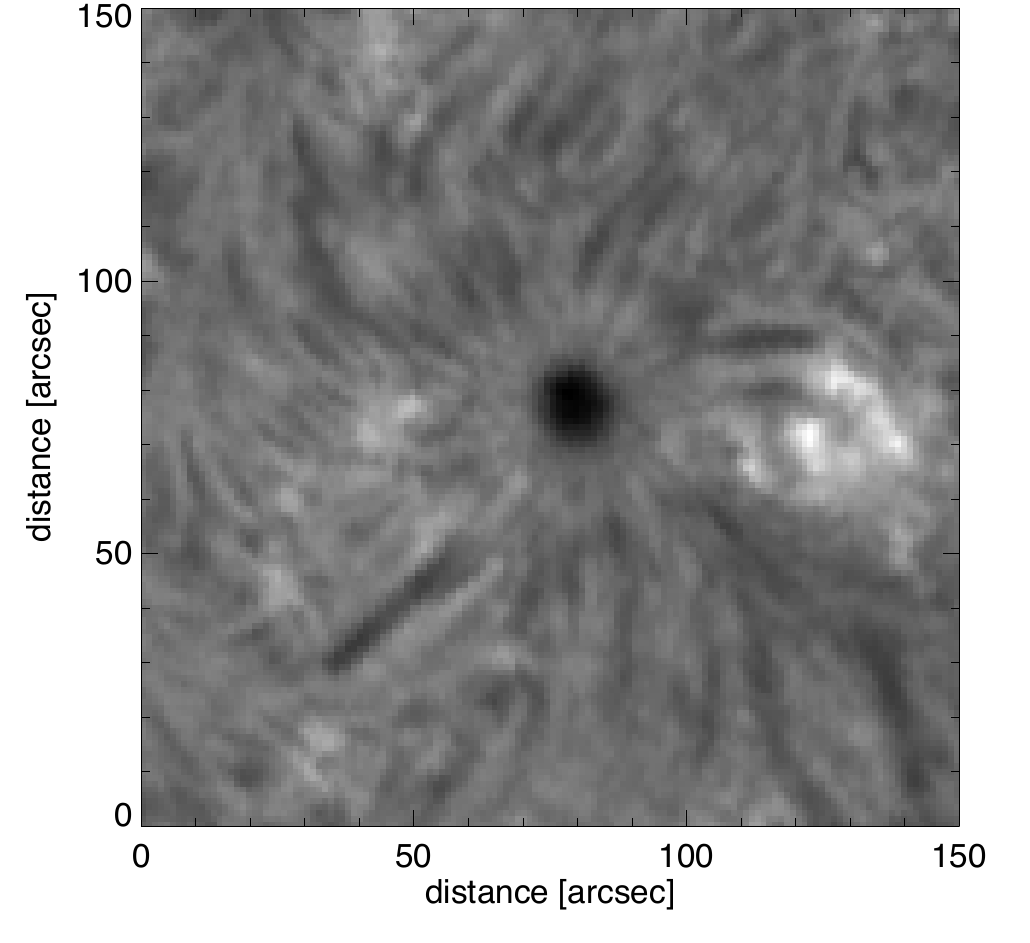}$\quad\,\,$\includegraphics[height=0.2851\textwidth]{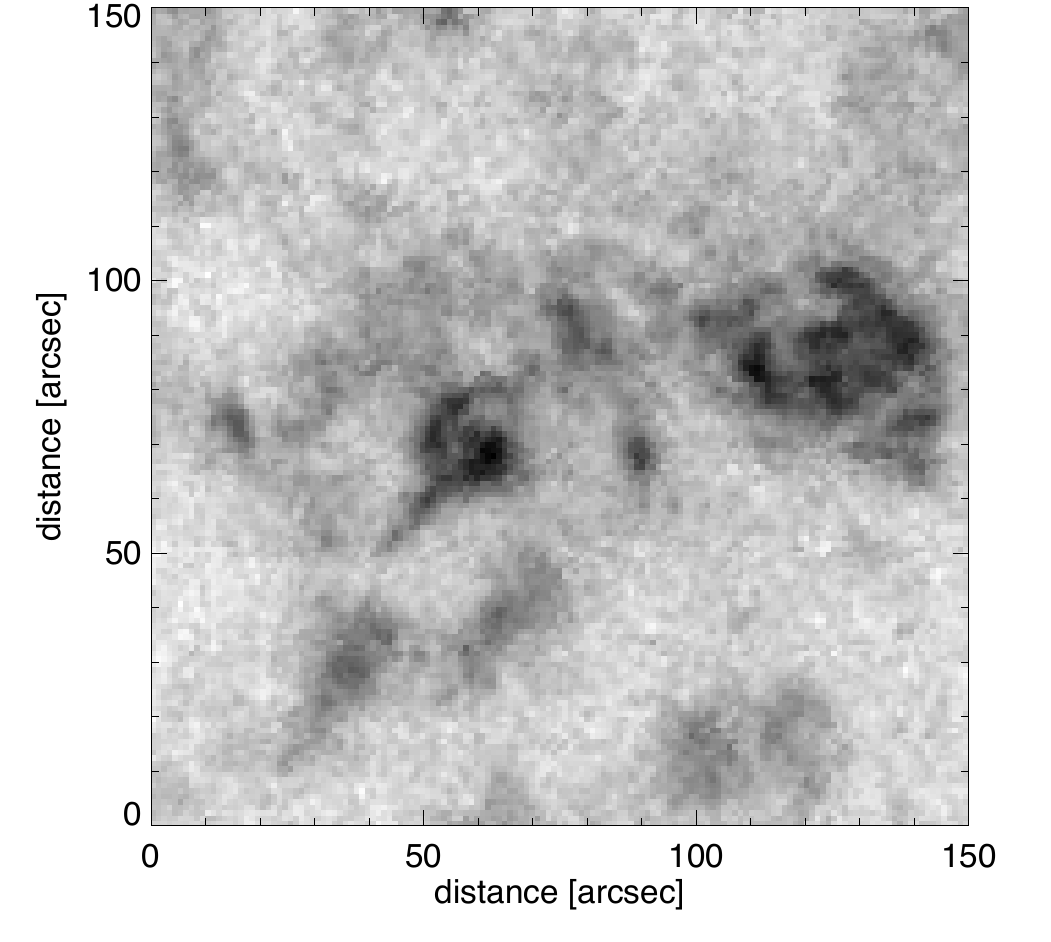}
  \caption{Sample images from ChroTel on 31 August 2010; full disk and
    full resolution cutouts. The black squares indicate a size of
    2 arcseconds in the bottom row. \emph{From left to right:} \ion{Ca}{ii} K
    (12:07:39 UTC), H$\alpha$ (10:37:16 UTC), and \mbox{\ion{He}{i}}
    10830\,\AA$\;$(11:57:17 UTC). An image in the nearby continuum was
    subtracted from the \mbox{\ion{He}{i}} image to enhance the contrast.}
  \label{fig:samples}
\end{figure*}

Figure~\ref{fig:samples} presents some ChroTel full-disk images in all
channels acquired on 31 August 2010. In addition, cutouts are shown to give a visual impression
of the full spatial resolution of the images. The bottom row
demonstrates that the smallest visible structures in H$\alpha$ and
\ion{He}{i} are close in size to the theoretical resolution limits of
2~arcsec in H$\alpha$ and 2.7~arcsec in \ion{He}{i}. In
\ion{Ca}{ii} K, the theoretical limit of 2~arcsec is not entirely
achieved owing to the long exposure time that is necessary
in this channel and the higher susceptibility to seeing at shorter
wavelengths.

The mean spatial resolution of the data was computed for 
three sample days in summer 2010 (July 26, July 31, August 3). In a
box of 800x800 pixels covering both quiet Sun and active regions, the
Fourier power as a function of spatial frequency was determined. A
flat part in this curve, i.e., when the Fourier power is the same for
all frequencies, represents the Fourier transformation of white noise
where no information is left in the data. The point where the curve
becomes flat therefore marks the resolution limit. The
determination of this point is not exact, but nevertheless gives a
rough estimate of the spatial resolution of the data. For the data of
the aforementioned three days, the mean spatial resolution in all
channels turned out to be about two times worse than the theoretical
resolution limit, i.e., the smallest resolved structures are about
twice as large. This is predominantly caused by deteriorating seeing
conditions in the afternoon.  

For the same days, the mean scattered light at
1.1$\cdot\,\rm{R_{\odot}}$ was determined as 0.13 in \ion{Ca}{ii} K
and 0.04 in both H$\alpha$ and \ion{He}{i} relative to the mean
intensity on the solar disk.  

\section{Line-of-sight velocities in \mbox{\ion{He}{i}} 10830\,\AA\label{sec:doppler}}
Figure~\ref{fig:he_pb} illustrates that a range of
$\pm$3\,\AA~around the \mbox{\ion{He}{i}} triplet is covered by the seven
filtergrams. Theoretically, this should enable us to detect LOS
velocities of up to about $\pm$80\,km\,s$^{-1}$. One possible way to 
determine the position $\lambda_{line}$ of the line in every pixel is a
center-of-gravity approach using the filtergram intensities $I_i$ and
the central wavelengths $\lambda_i$ of the filter passbands known from
measurements
 
\begin{equation}\label{eqn:cog}
 \lambda_{line}=\frac{\sum_i(1-I_{i})\lambda_{i}}{\sum_i(1-I_{i})}.
\end{equation}

\noindent Strong absorption in a filtergram therefore leads to a
strong weighting of its central wavelength. Equation (\ref{eqn:cog}) 
is applicable for an undisturbed single spectral line and
filtergrams that are normalized to the continuum value. For real data
however, a straightforward determination of the line-shift is affected
by the influence of other spectral lines, different continuum levels,
different line widths and depths, a varying ratio of the blue to red
part of the \mbox{\ion{He}{i}} triplet, multiple velocity components in a
resolution element, and instrumental effects. The theoretical approach described in
Eq.~(\ref{eqn:cog}) therefore has to be extended to account for these
(partially unknown) influences. This was addressed by calibrating the
line-shift maps generated from the filtergrams with measured
line-shift maps from a spectrograph. 

In December 2007, parallel observations in \mbox{\ion{He}{i}} 10830\,\AA~were made
with ChroTel and the Tenerife Infrared Polarimeter
\citep[TIP-II,][]{Collados2007}. The TIP-II
scans covered three sections of the active region NOAA 10978 on 8 December
2007, each of size 82x100 arcsec. The stepsize in the
scanning direction was 0.5\,arcsec; each scan took 15\,minutes
(UTC 11:05:28-11:20:58, 11:23:59-11:39:31, 12:03:31-12:19:07). In parallel, ChroTel
observations were performed solely in the \mbox{\ion{He}{i}} channel with the
highest possible cadence of 30\,s. 

Line-shift maps were compiled from the TIP-II scans by
determining the position of the minimum intensity of a smoothed spectrum in the
wavelength range between 10828.5 and 10831.9\,\AA~in every pixel,
thus between the photospheric \ion{Si}{i} line and the water vapor line
indicated in Fig.~\ref{fig:he_pb}. This was done with a precision of
one pixel in the spectra, which corresponds to a precision in velocity of
{\raise.17ex\hbox{$\scriptstyle\sim$}}300\,m\,s$^{-1}$. The maps exhibit a large range of velocities,
i.e., blueshifts up to $-$14~km\,s$^{-1}$ and redshifts up to 43~km\,s$^{-1}$.

The regions scanned with TIP-II were cut out from the ChroTel full-disk
filtergrams and carefully aligned with the \mbox{TIP-II} maps. To simulate the
scanning procedure and the temporal evolution of the
observed structures during the scanning procedure, we first tried
to take only stripes from the ChroTel filtergrams closest in time to
the actual slit step and to compose an artificial filtergram set from
these stripes \citep[cf.][Appendix B]{Beck2007}. The varying seeing
conditions in the ChroTel images 
however prohibited a reliable composition. We therefore decided to
take only the exposure with the best seeing conditions during the
15~minutes of the TIP-II scan (at UTC 11:15:52, 11:25:15, and 12:07:08). 
This is justified by the steadiness of most structures seen in the
scans, which was verified using movies compiled from the ChroTel images.   

\begin{figure*}
\sidecaption
 \includegraphics[width=12cm]{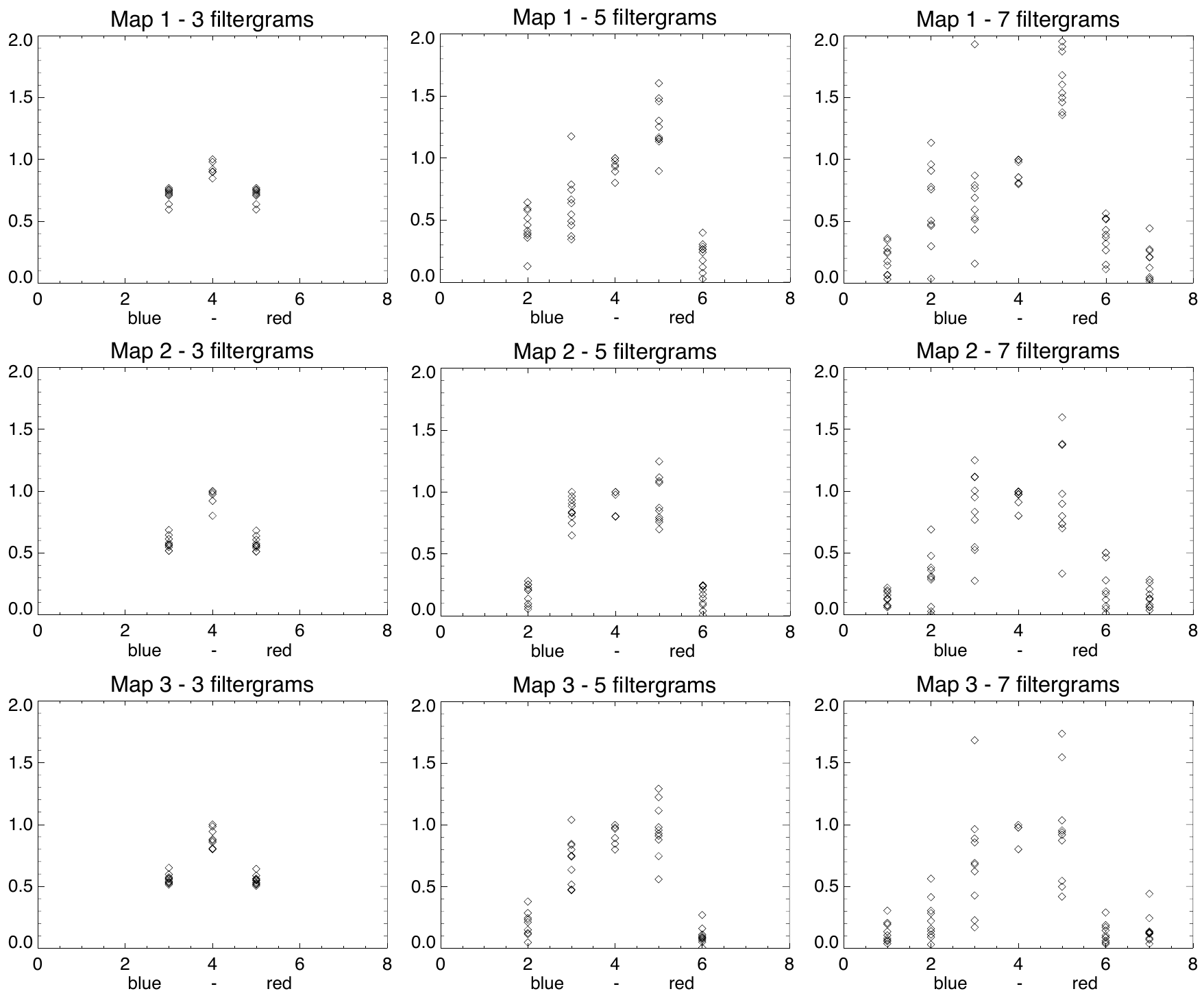}
  \caption{$\alpha_i$ factors for the reconstruction of the three
    velocity maps with a different number of used filtergrams. The
    abscissa denotes the number of the filtergram according to
    Fig.~\ref{fig:he_pb}. Each run was repeated ten times to see whether
    the $\alpha_i$ converge to a single set (see Sect.~\ref{sec:convergence}).} 
  \label{fig:parameter_statist}
\end{figure*}

To perform a first order correction of the possibly varying influence of
the prefilter transmission curve on the measured intensities at
different wavelengths, the median filtergram intensities
$\overline{I_i}$ were normalized to the median intensity in the second
filtergram. The latter has the least contribution from any spectral line
and therefore comes closest to a continuum value. The normalization
was done by computing the median value in a centered square covering
about 37\% of the solar disk in each filtergram. Each filtergram was
subsequently multiplied by the ratio of the median intensity of the
second to the actual filtergram

\begin{equation}\label{eqn:itilde}
 \tilde I_i=\frac{\overline{I_2}}{\overline{I_i}}\,I_i .
\end{equation}

\noindent With the obtained intensities $\tilde I_i$, the line-shift maps were then
computed based on Eq.\,(\ref{eqn:cog}) according to

\begin{equation}\label{eqn:cogn}
 \Delta\lambda_{recon}=\frac{\sum_i(I_c-\alpha_{i}\tilde
   I_{i})\lambda_{i}}{\sum_i(I_c-\alpha_{i}\tilde I_{i})}-\lambda_0 , 
\end{equation}

\noindent where $\lambda_0$ is the rest wavelength of the \mbox{\ion{He}{i}} line at
10830.3\,\AA, and $I_c$ represents the continuum intensity in every pixel,
which was estimated to be 

\begin{equation}\label{eqn:cont}
 I_c=\frac{I_2}{0.95} .
\end{equation}

\noindent The $\alpha_i$ in Eq.~(\ref{eqn:cogn}) are arbitrary factors that
assign an enhanced or decreased weighting to a specific filtergram,
hence to a specific wavelength. A variation in these factors thus 
leads to a different result for the calculated line-shift map. A
specific set of $\alpha_i$ can be considered optimal when the sum over
all pixels of the squared difference of the real line-shifts
$\Delta\lambda_{real}$ computed from the TIP-II scans and the
reconstructed line-shifts $\Delta\lambda_{recon}$ from the filtergrams
is smallest, i.e., when 

\begin{equation}\label{eqn:sum}
 \sum_{pixels}(\Delta\lambda_{real}-\Delta\lambda_{recon})^2
\end{equation}

\noindent is minimal or

\begin{equation}\label{eqn:sumv}
 \sum_{pixels}(v_{real}-v_{recon})^2
\end{equation}

\noindent when the line-shifts are expressed as Doppler velocities according to
$v=(\Delta\lambda/\lambda_0)\cdot c$. 

\begin{figure*}
\includegraphics[width=0.02\textwidth]{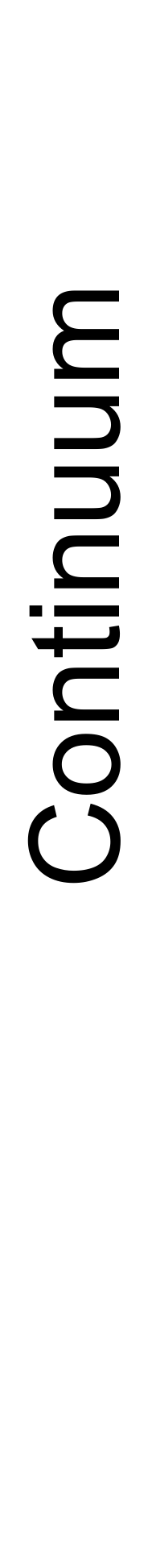}$\;$\includegraphics[width=0.27205\textwidth]{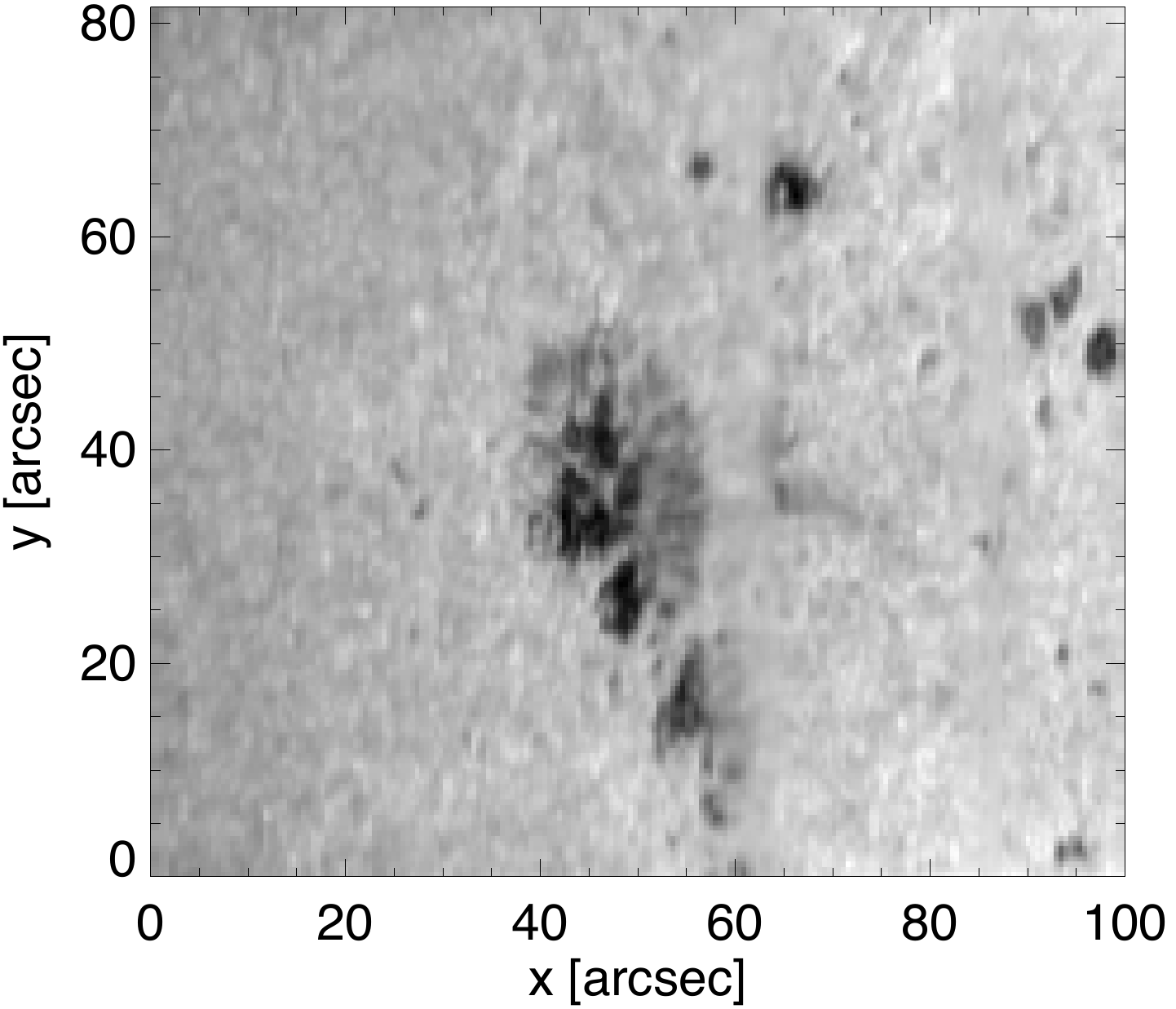}\hspace{29.2pt}\includegraphics[width=0.27205\textwidth]{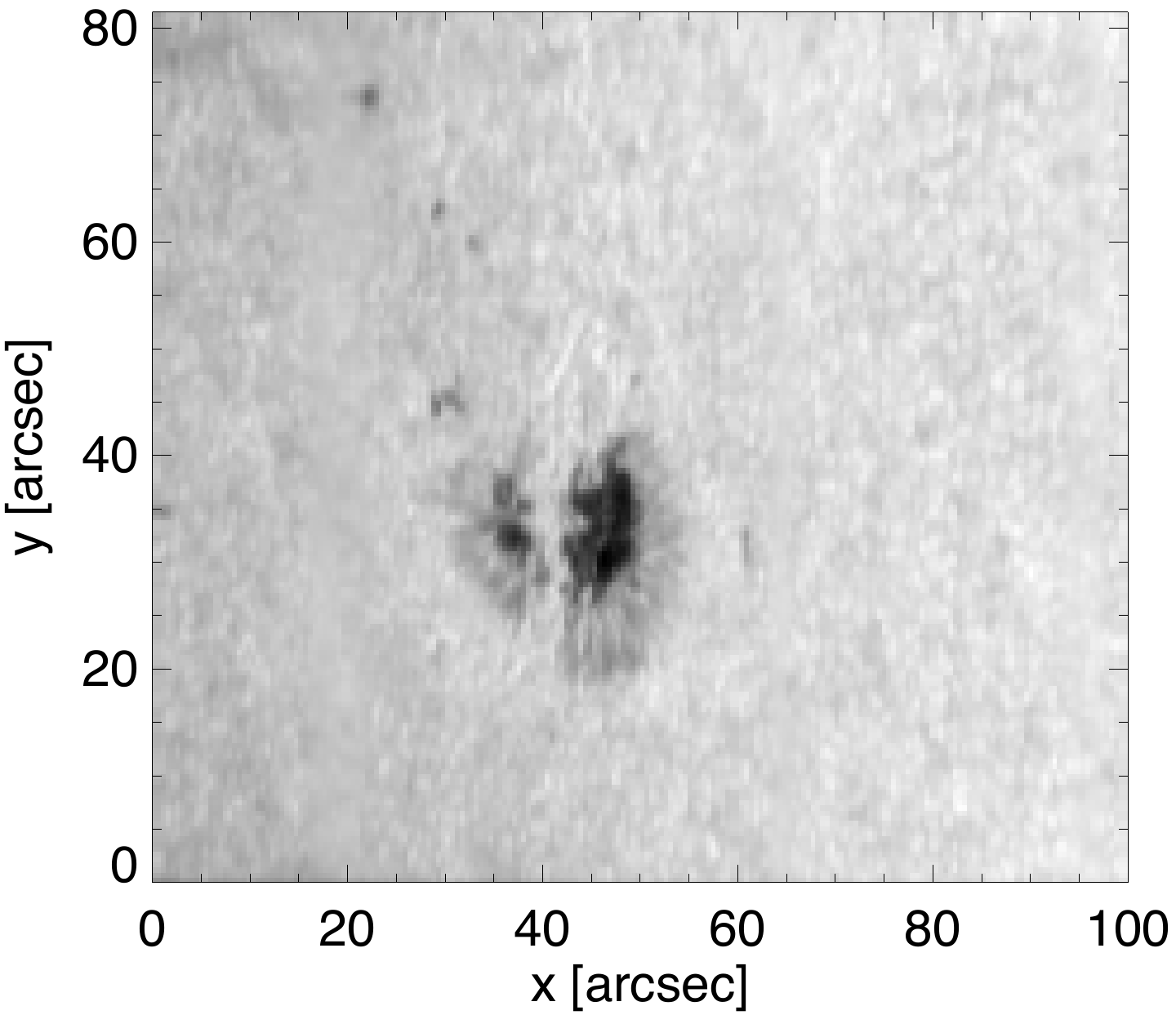}\hspace{29.6pt}\includegraphics[width=0.273\textwidth]{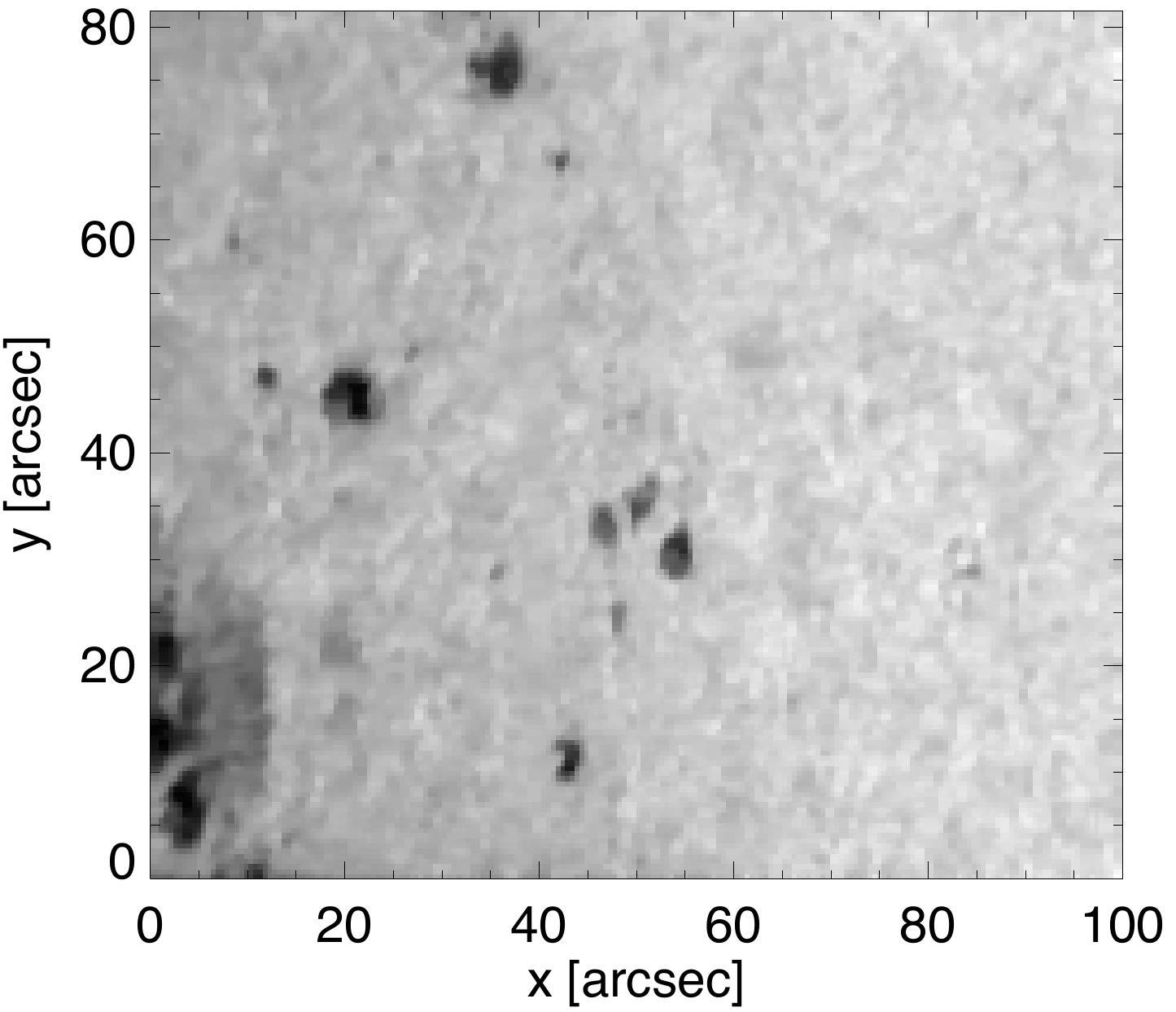}\\[0.15cm]\includegraphics[width=0.02\textwidth]{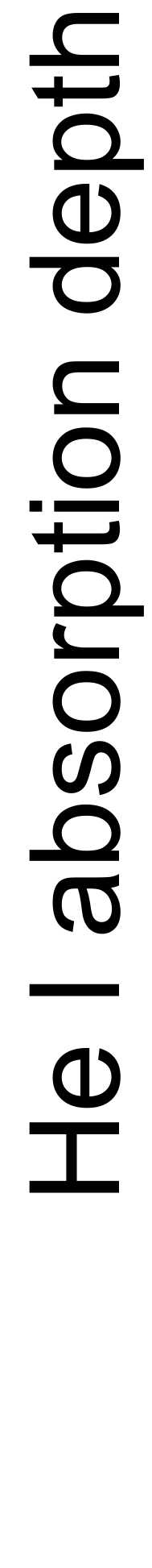}$\;$\includegraphics[width=0.313\textwidth]{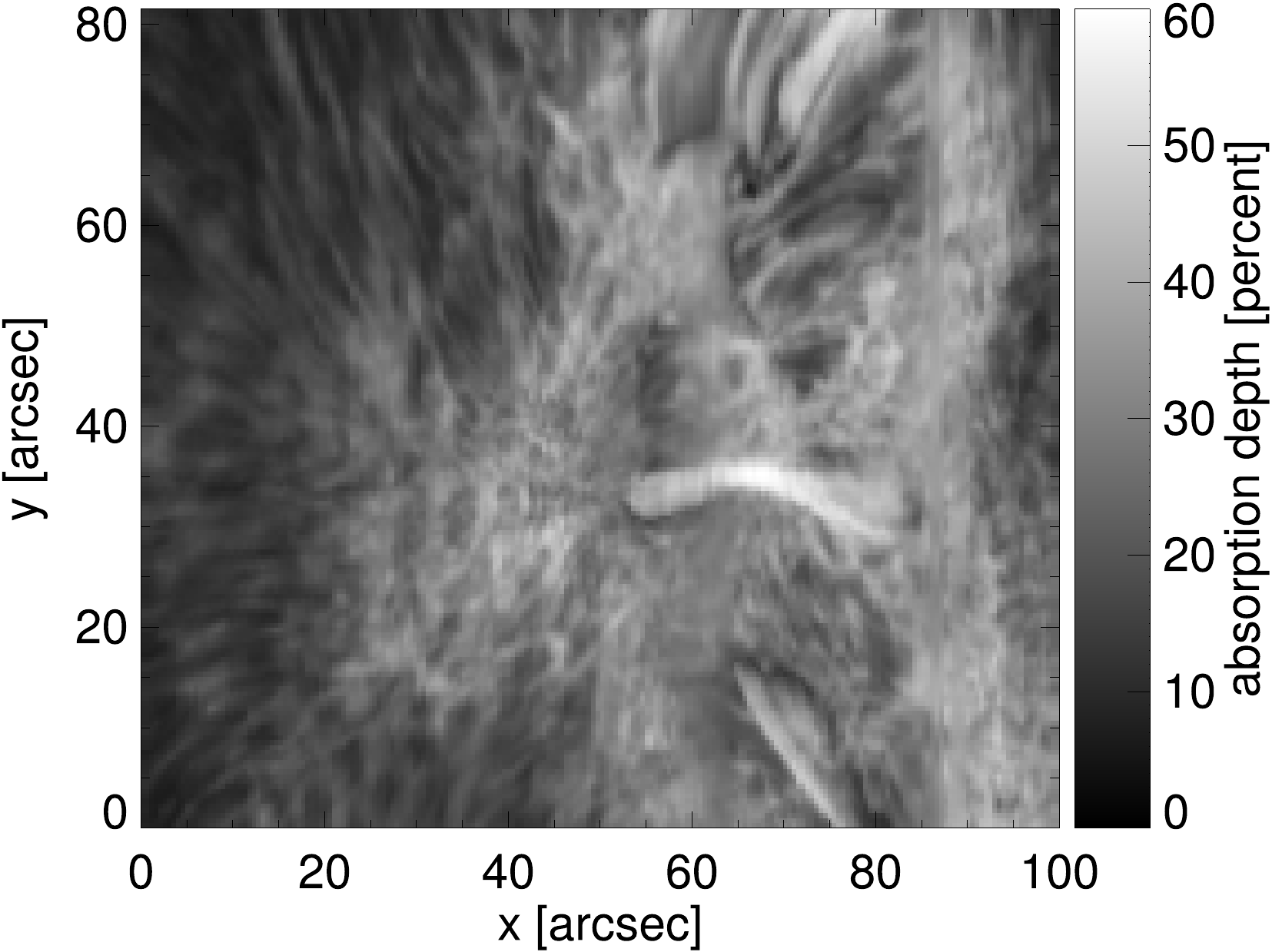}$\;\;\,\,$\includegraphics[width=0.313\textwidth]{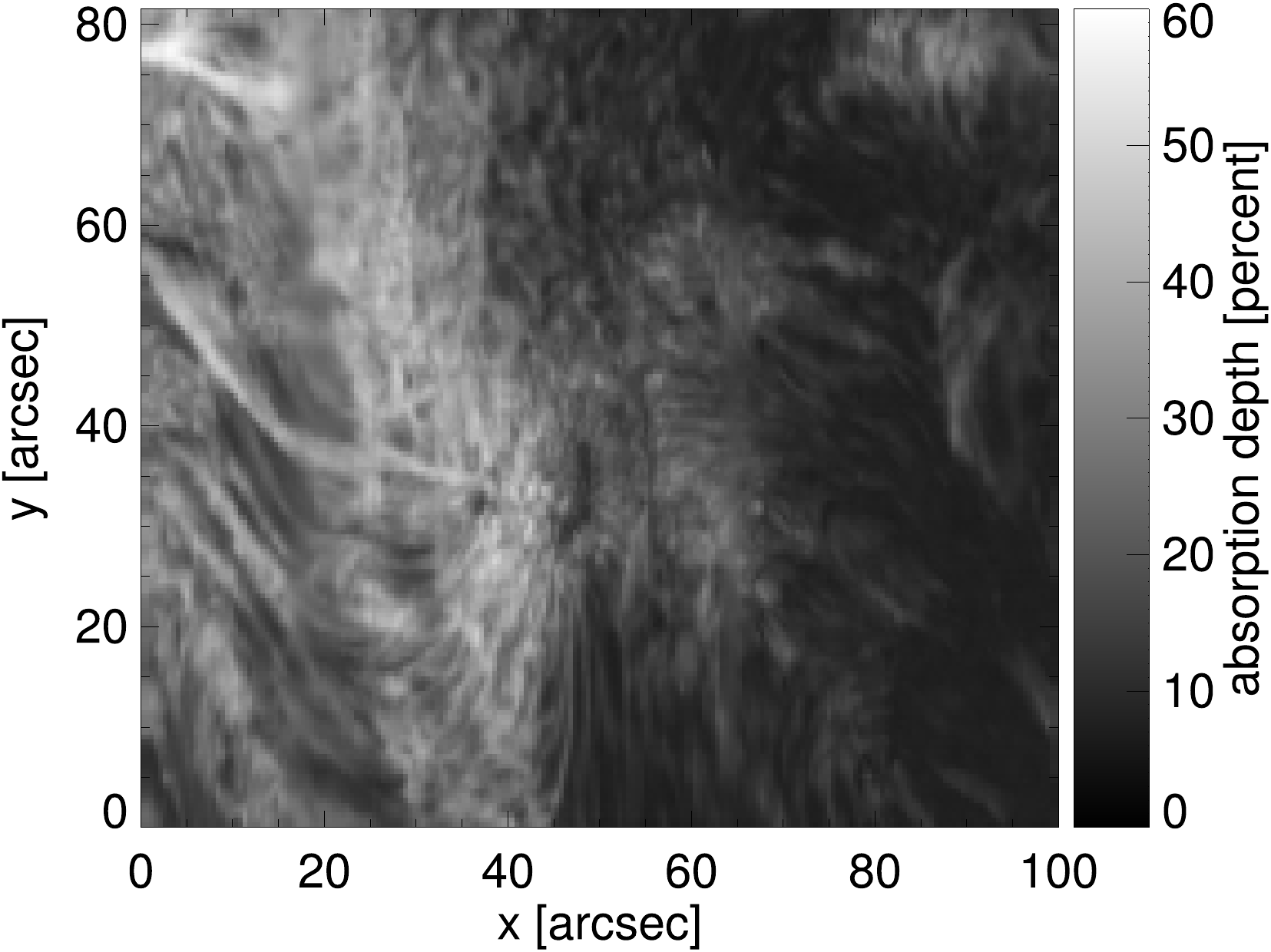}$\;\;\,\,$\includegraphics[width=0.313\textwidth]{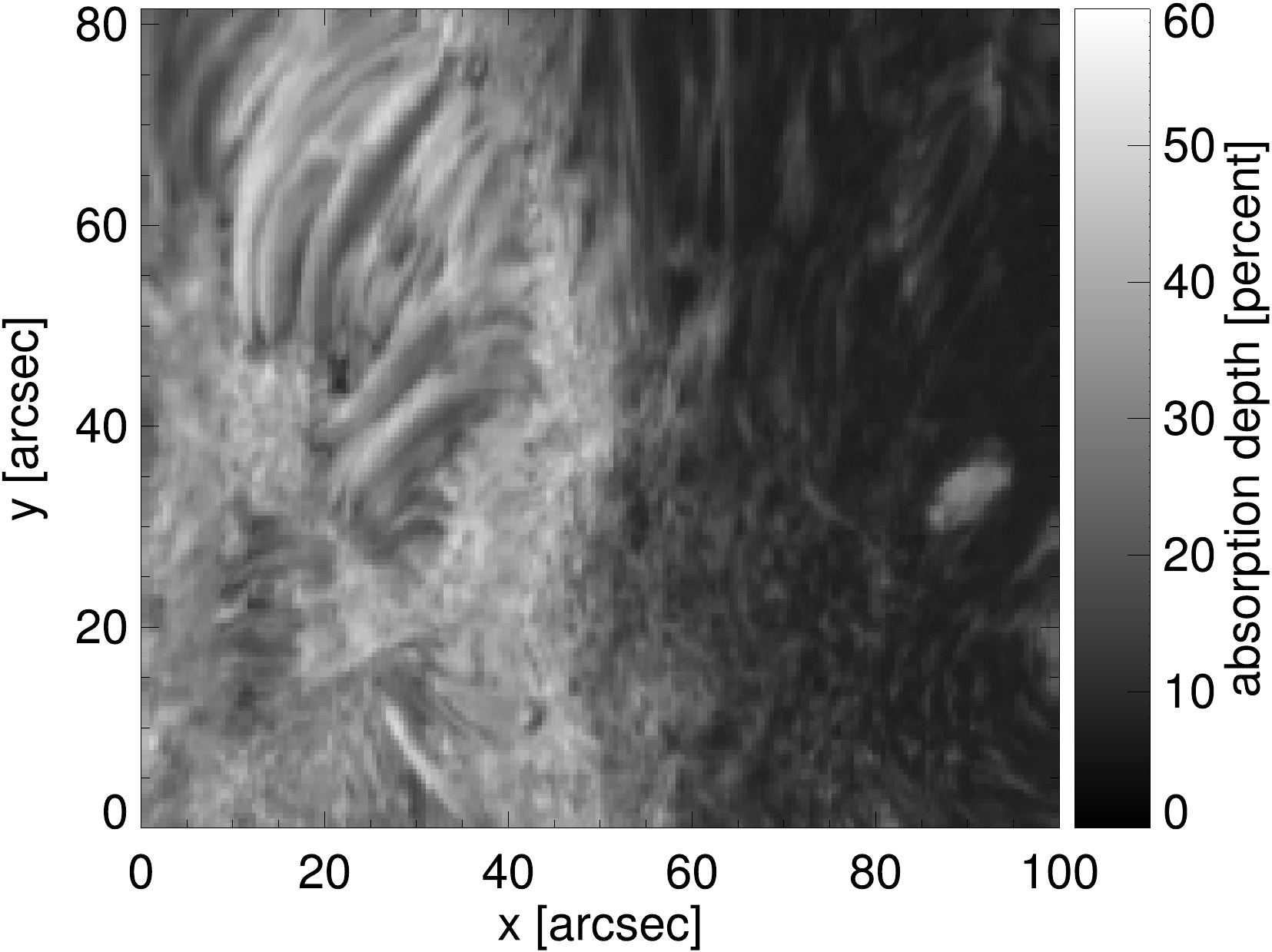}\\[0.15cm]\includegraphics[width=0.02\textwidth]{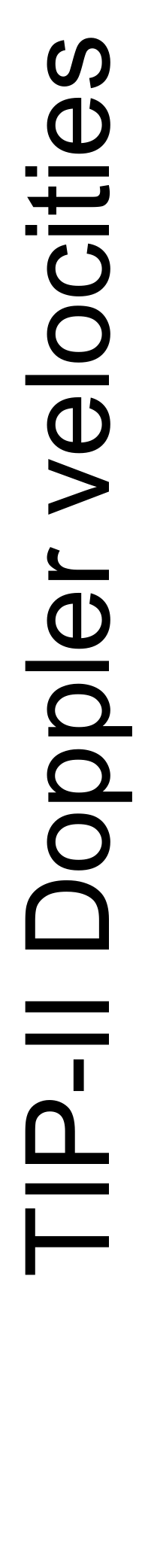}$\;$\includegraphics[width=0.316\textwidth]{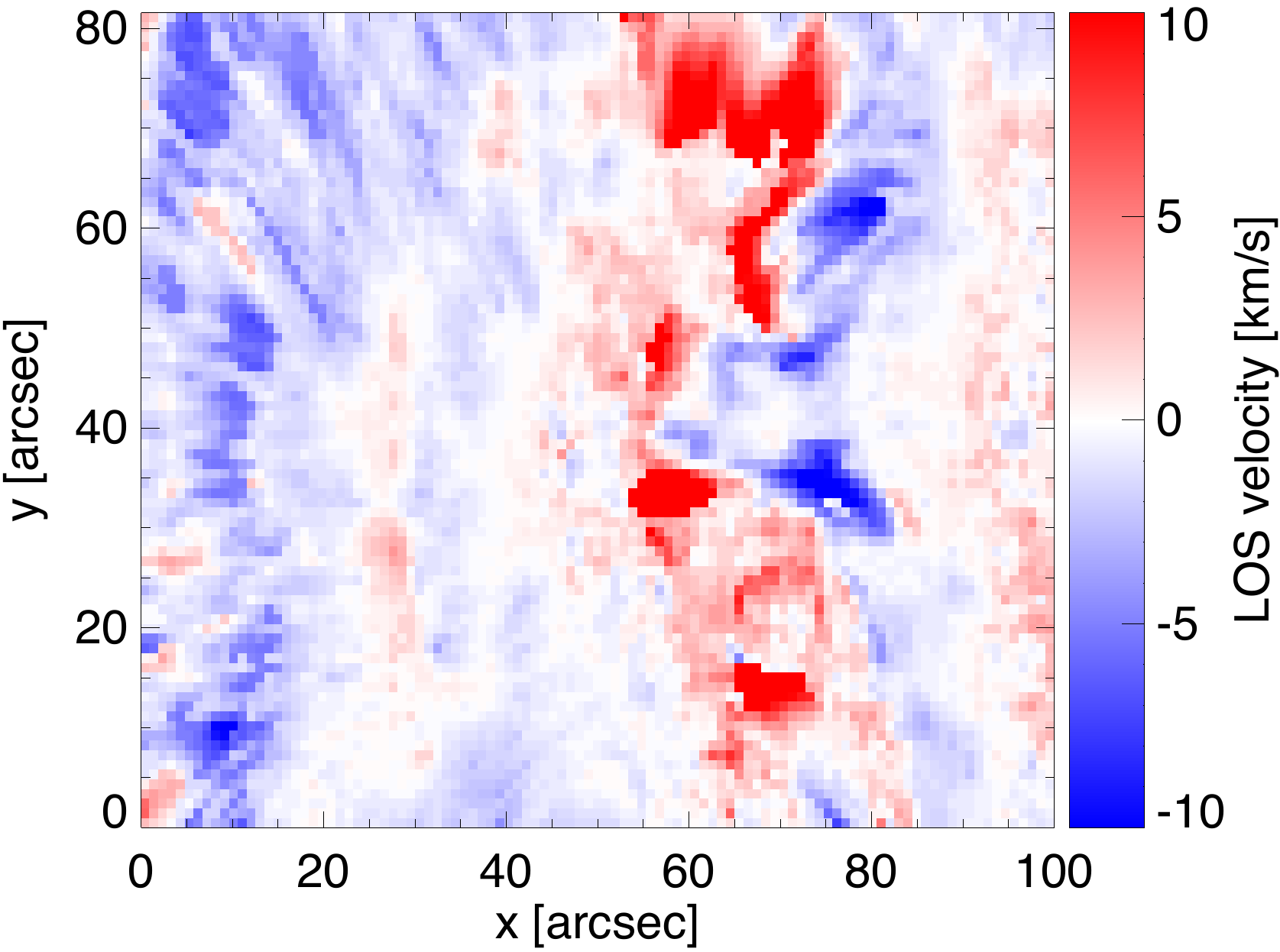}$\;\;\,$\includegraphics[width=0.316\textwidth]{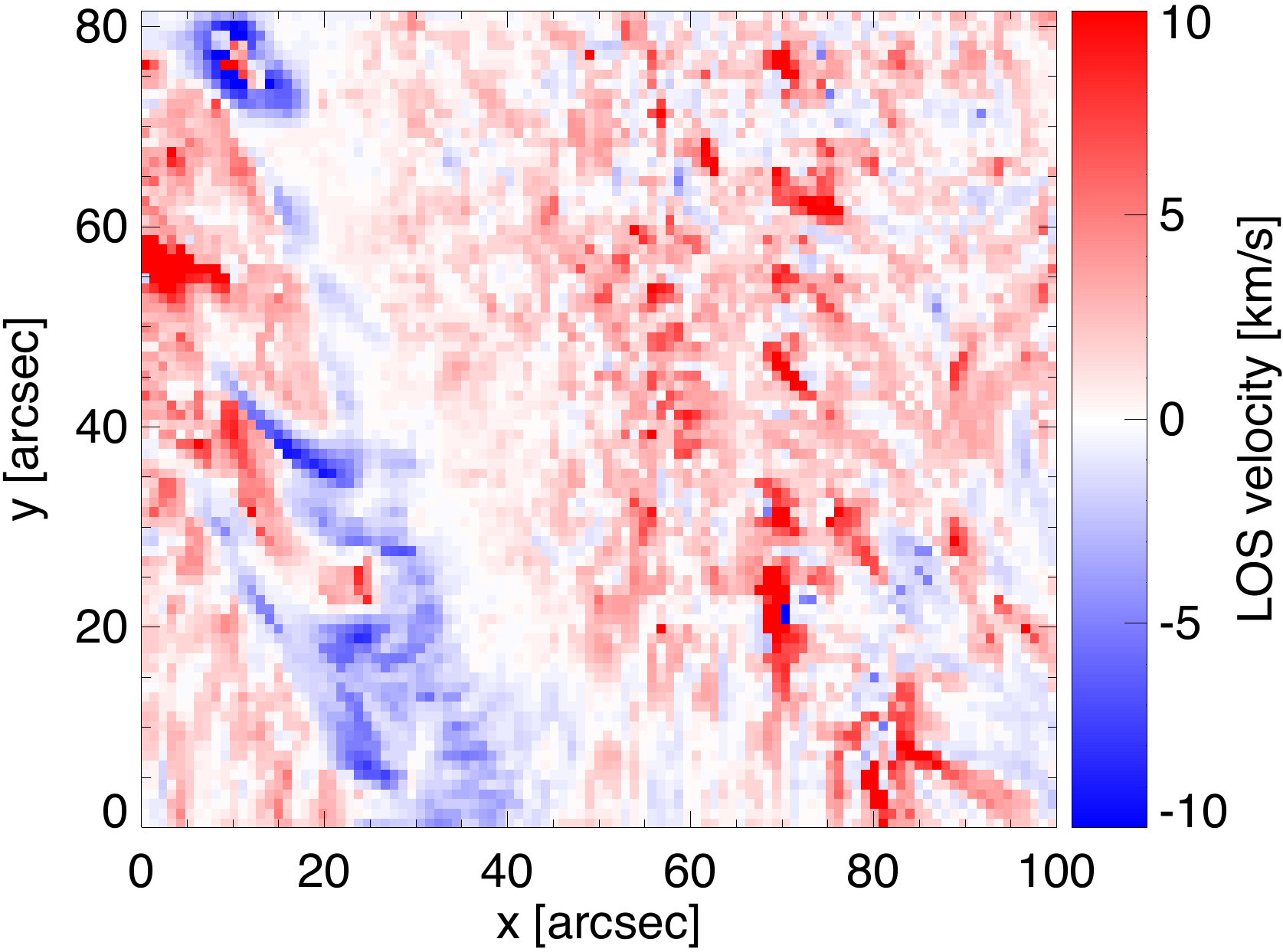}$\;\;\,$\includegraphics[width=0.316\textwidth]{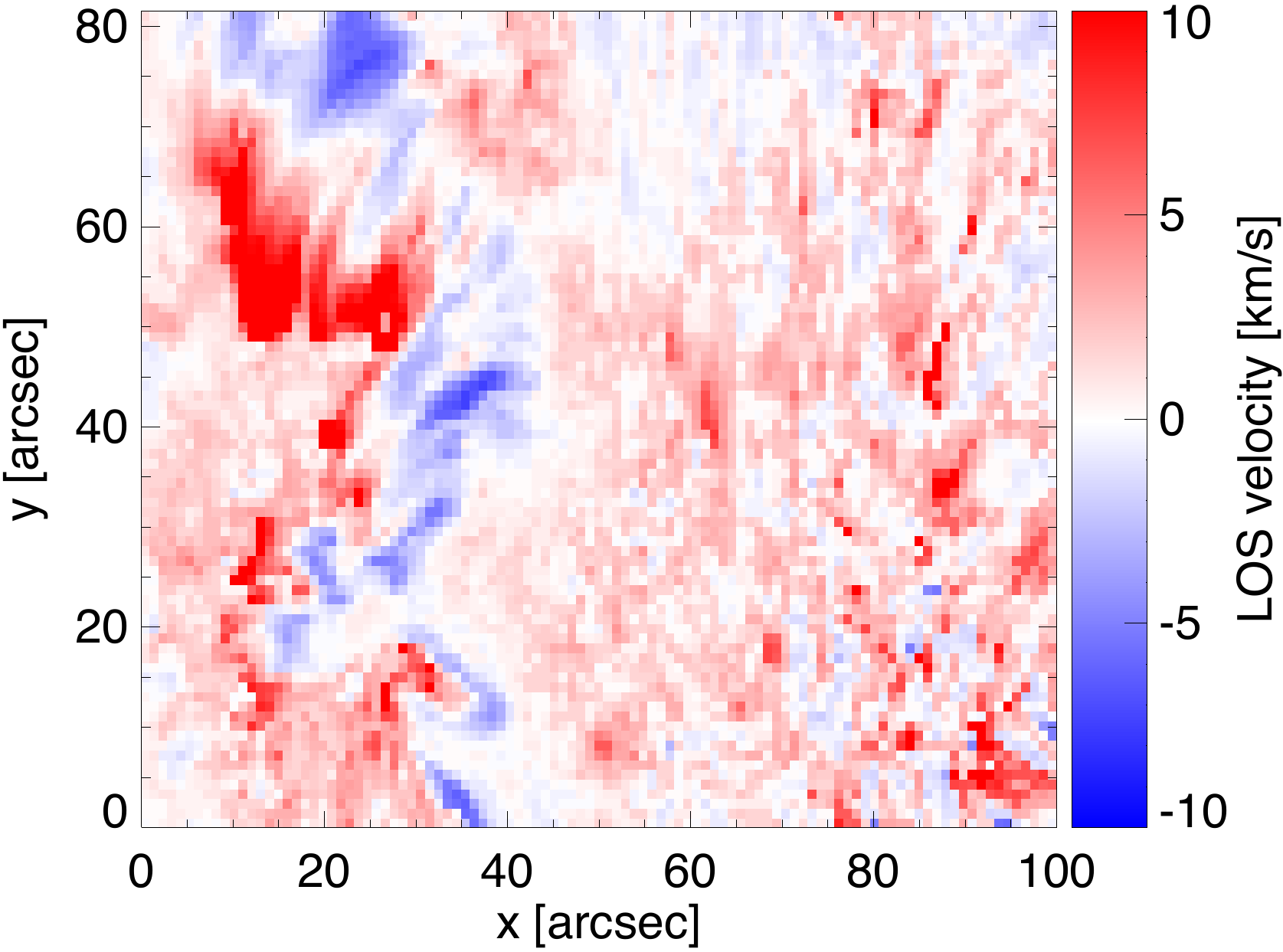}\\[0.15cm]\includegraphics[width=0.02\textwidth]{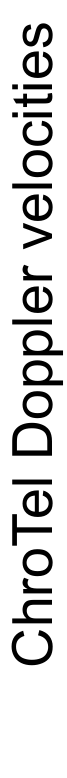}$\;$\includegraphics[width=0.316\textwidth]{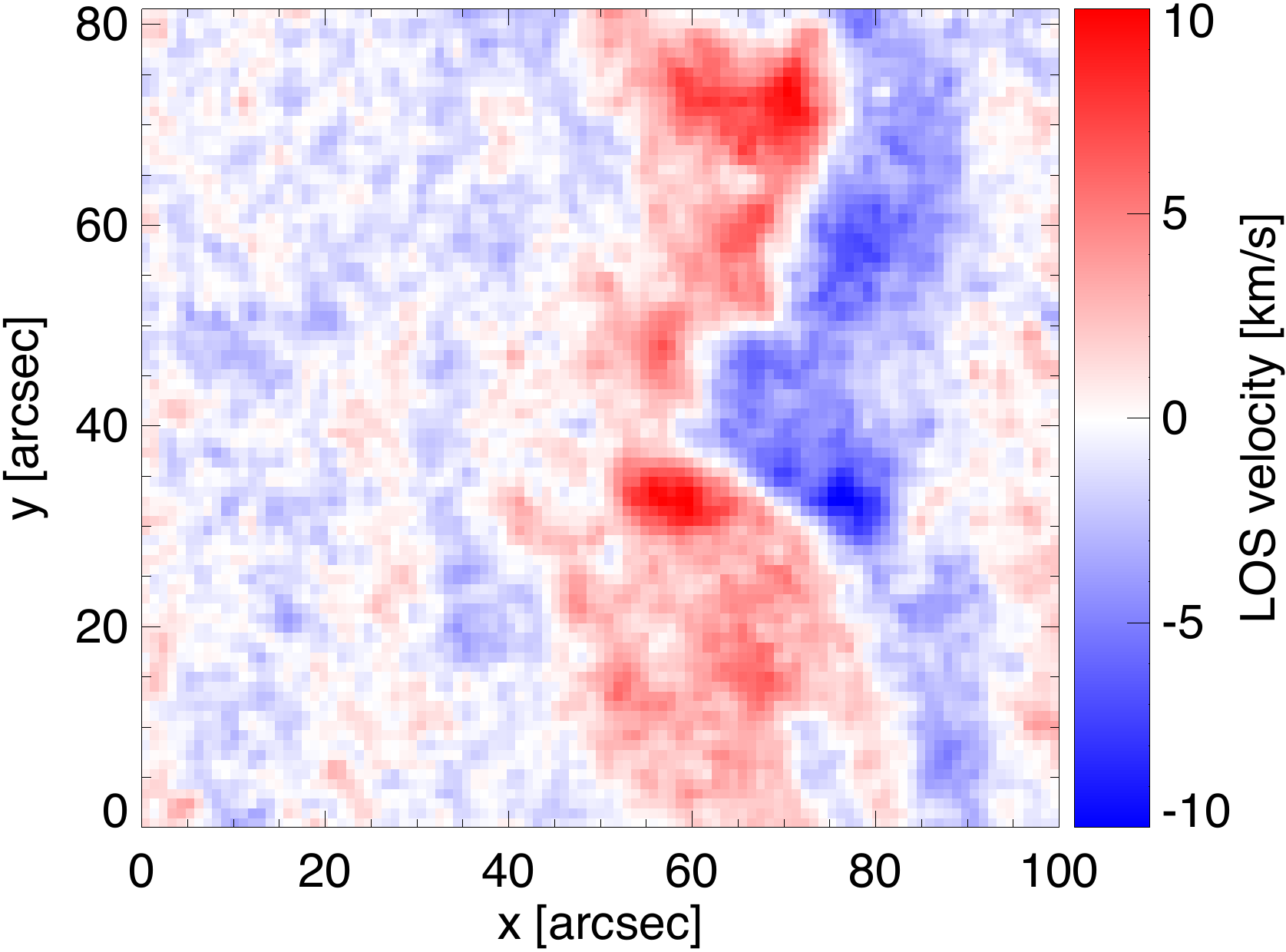}$\;\;\,$\includegraphics[width=0.316\textwidth]{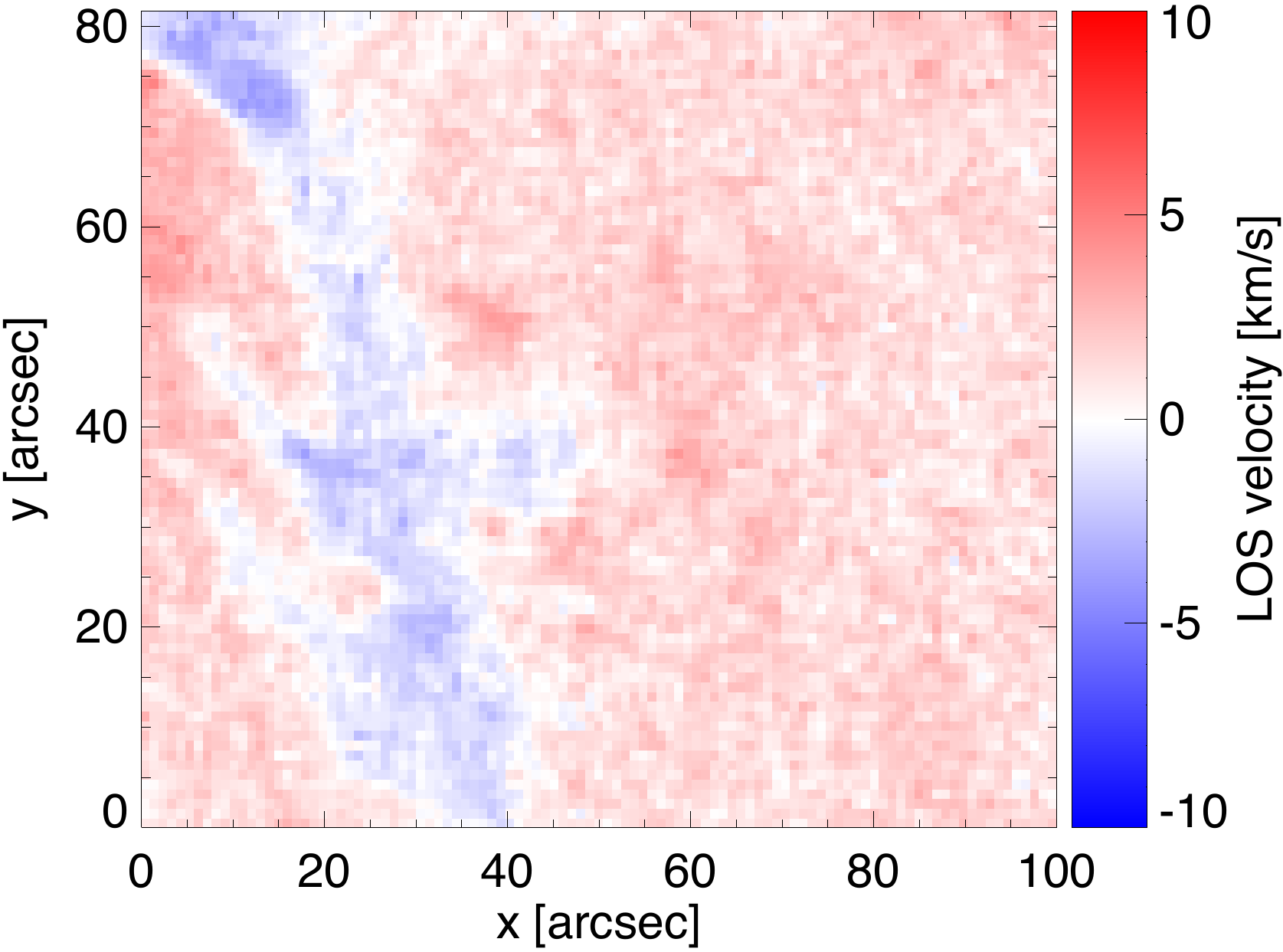}$\;\;\,$\includegraphics[width=0.316\textwidth]{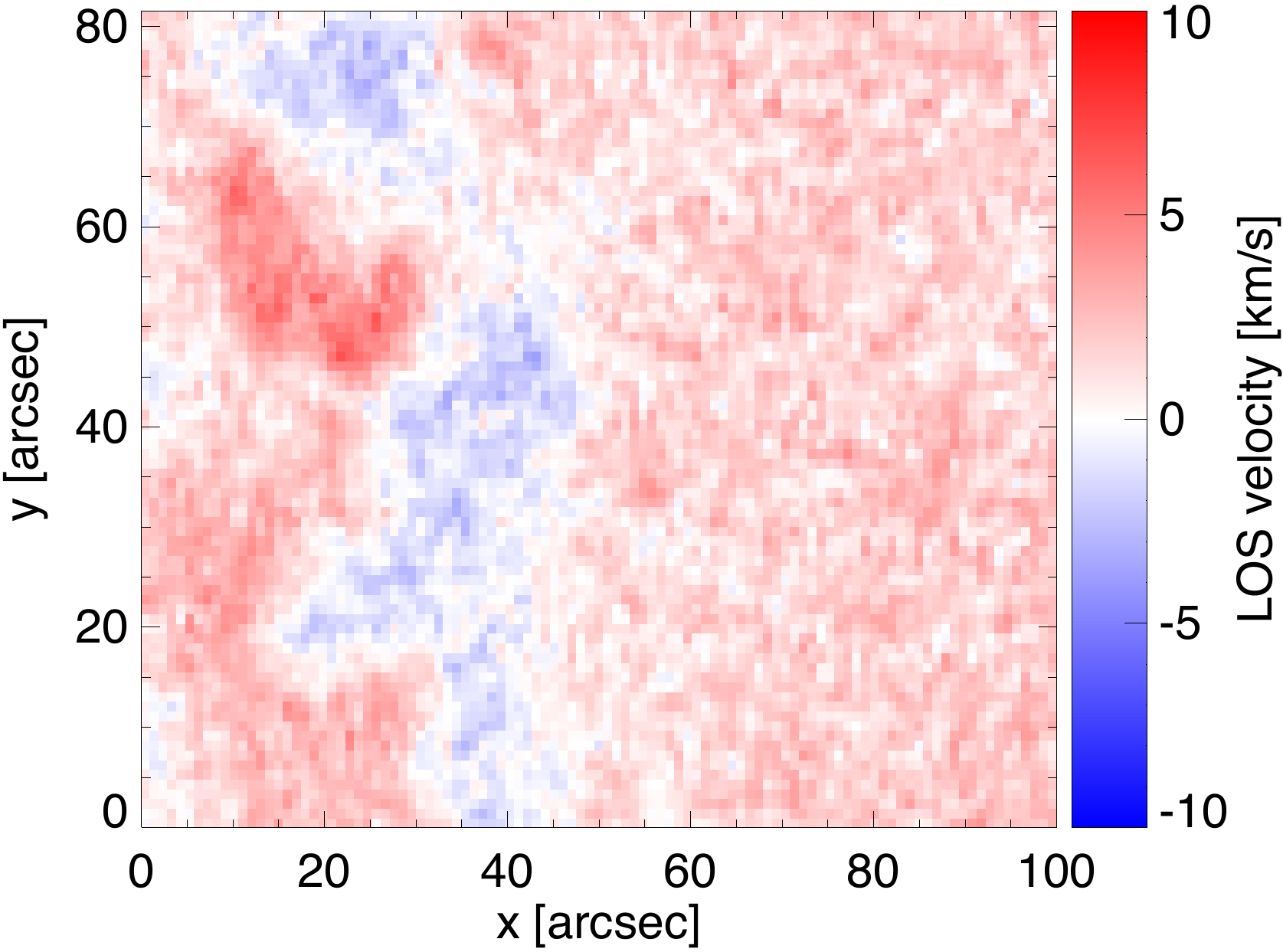}
  \caption{\emph{Rows from top to bottom:} Continuum maps of the three TIP-II
    scans used for the velocity calibration for ChroTel. \ion{He}{i}
    10830\,\AA$\;$absorption depth. \ion{He}{i} LOS velocities as determined from the
    TIP-II  spectra. \ion{He}{i} LOS velocities as reconstructed from the
    ChroTel filtergrams (only filtergrams \#3-\#5; one set of
    $\alpha_i$, same weighting for all pixels. See Sect.~\ref{sec:calib_runs}.).} 
  \label{fig:velo_maps}
\end{figure*}

Finding this set of $\alpha_i$ is an optimization problem that was
addressed with the genetic algorithm PIKAIA
\citep{Charbonneau1995}. Genetic algorithms are comparably slow, but
are known to find global minima more reliably because they employ a random
mutation of the fit parameters, which allows them to escape from local minima.

\subsection{Calibration runs\label{sec:calib_runs}}
Calibration runs were conducted with all three TIP-II scans using three, five,
and seven filtergrams for the reconstruction. If less than seven
filtergrams were used, those that are closest to line center were
chosen, e.g., \#2\,-\,\#6 in Fig.~\ref{fig:he_pb} for five filtergrams. The
runs using a different number of filtergrams were intended to show whether
the additional information from the outer filtergrams improves the fit
results or might be obsolete.

The values of the $\alpha_i$ were allowed to vary between 0 and 2. 
By default, every pixel was treated the same. In additional runs,
enhanced weightings were applied to pixels with high absorption depth
and to pixels showing small velocities because the latter
contribute the largest fraction. Each run was repeated ten
times to see whether the results of the optimization algorithm
(ideally) converge to a single set of $\alpha_i$. The $\alpha_i$ from the ten runs and
from all three maps were then averaged to compile velocity maps for a
quantitative analysis. 

\subsection{Convergence of the $\alpha_i$\label{sec:convergence}}
Figure~\ref{fig:parameter_statist} shows the $\alpha_i$ parameters for
all three calibration maps applying a different number of filtergrams
for ten runs each. The distribution of the parameters is
consistent for the three maps within a fixed number of used
filtergrams. \mbox{Map 1} shows slightly enhanced weightings of filtergram 5
and decreased weightings of filtergram 3 for the use of five and seven
filtergrams, which is because of the high redshifts occurring in this
map. For three filtergrams, the code always applies almost the same
weighting to filtergram 3 and 5, so the velocities are
apparently not reflected in the ratio of $\alpha_5$ to $\alpha_3$, but
in the ratio of both $\alpha_3$ and $\alpha_5$ to $\alpha_4$. 

As the number of filtergrams used increases, the spread in the
$\alpha_i$ parameters also increases for the ten runs, indicating that the
fits become more arbitrary as the number of fit parameters increases. While
the $\alpha_i$ are mostly confined to within a range of 0.5 for five
filtergrams, we find a spread over almost the whole range for seven
parameters. This indicates that the application of only three filtergrams
seems to be the best choice as long as the velocities are not too large.

\subsection{Calibration results\label{sec:calib_results}}
\begin{figure*}
 \sidecaption
  \includegraphics[width=12cm]{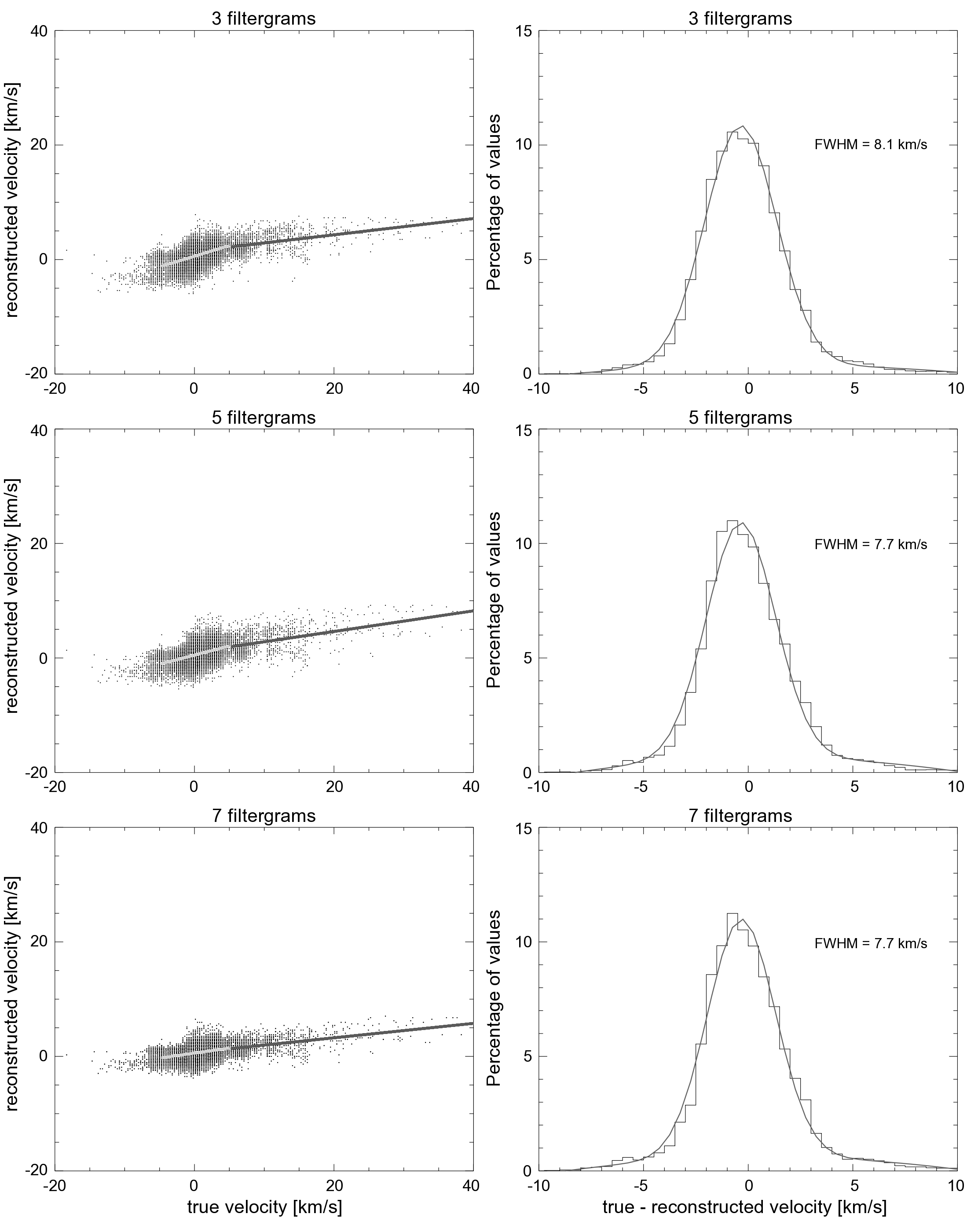}
  \caption{\emph{Left column:} True vs. reconstructed velocities for
    the use of three, five, and seven filtergrams. The light grey and dark grey lines
    depict linear-regression fits to the velocity regimes
    \mbox{\,$-$5~$< v <$~5~km\,s$^{-1}$} $\;$and$\;$ \mbox{$v
      \geqslant$~5~km\,s$^{-1}$}, respectively. \emph{Right column:} Histograms of the
    velocity difference for the true and reconstructed velocities. The
  FWHM is calculated from a Gaussian fit to the histograms (see Sect.~\ref{sec:nof}).}
  \label{fig:scatter_hist}
\end{figure*}

The two top rows in Fig.~\ref{fig:velo_maps} show the absorption depth
in \mbox{\ion{He}{i}} and the intensity in the nearby continuum of the
three TIP-II scans that were used for the velocity calibration. In the third 
row, the LOS velocities are shown as determined from the TIP-II
spectra. For each of these velocity maps, an example of a
corresponding velocity map reconstructed from the three innermost
ChroTel filtergrams with one set of $\alpha_i$ is shown for comparison
in the lowermost row. 

As can be expected from both the lower spatial and spectral
resolution, the ChroTel velocity maps lack the fine structures exhibited in the TIP-II
maps. This seems to be the case primarily in regions of weak
\mbox{\ion{He}{i}} absorption, as indicated by the two blueshifted streaks in
the second map at about x~=~15$''$. In contrast to the fine
structures on the right-hand side of the map (where the \mbox{\ion{He}{i}}
absorption is low), the two blueshifted streaks are clearly visible in
the ChroTel velocity map, although they are not larger in size. For
very low absorption depths, the spectral resolution of the ChroTel
filtergrams does not appear to provide the sensitivity to capture the location of
the helium-line core accurately, even if it can still be determined in the
high-resolution TIP-II spectra.

The overall velocity structure is recovered remarkably
well in regions of high absorption depth, although the reconstructed
velocities significantly underestimate the real values. This is confirmed in
Fig.~\ref{fig:scatter_hist}, where the left column shows a scatter
plot of the true versus the reconstructed velocities for the use of three, five,
and seven filtergrams. While the true velocities reach values of 40~km\,s$^{-1}$
and larger, the values of the reconstructed velocities never exceed
10~km\,s$^{-1}$.
   
The scatter plots should ideally display a linear relation over the
whole velocity range. We can indeed separate the velo\-city distribution
into two distinct ranges: \mbox{$-$5~$< v <$~5~km\,s$^{-1}$}  
and \mbox{$v \geqslant$~5~km\,s$^{-1}$}. For these regimes, the
Pearson's correlation coefficient was computed to see how closely the
distribution follows a linear relation (cf.~Table~\ref{table:pcc}). In
addition, histograms of the difference between the true and the
reconstructed velocities were computed and fitted with a Gaussian
profile (right column in Fig.~\ref{fig:scatter_hist}). 

\begin{table}
\center
\caption{Correlation coefficients for the Doppler shift calibration (see Sect.~\ref{sec:calib_results}).\label{table:pcc}}
\begin{tabular}{cccc}
  \hline\hline
  calibration & number of   & \multicolumn{2}{c}{\quad\, correlation coefficient} \\
  run         & filtergrams & $-5 < v < 5$~km\,s$^{-1}$ & $v \geqslant 5$~km\,s$^{-1}$\\[2pt]
  \hline\\[-7pt]
  \multirow{3}{*}{regular} & 3 & 0.44 & 0.47\\
  & 5 & 0.39 & 0.49\\
  & 7 & 0.30 & 0.49\\[2pt]
  \hline\\[-7pt]
  \multirow{3}{*}{small velocities} & 3 & 0.44 & 0.47\\
  & 5 & 0.39 & 0.49\\
  & 7 & 0.30 & 0.50\\[2pt]
  \hline\\[-7pt]
  \multirow{3}{*}{high abs. depth} & 3 & 0.44 & 0.47\\
  & 5 & 0.38 & 0.49\\
  & 7 & 0.29 & 0.49\\[2pt]\hline
\end{tabular}
\end{table} 

\subsubsection{Number of filtergrams\label{sec:nof}}
The FWHMs of the histograms in Fig.~\ref{fig:scatter_hist} suggest
that the calibration gets successively better with the number of
filtergrams used. While this seems to be confirmed by the
correlation coefficients for redshifts larger than 5~km\,s$^{-1}$, the
opposite is seen for velocities of \mbox{$-$5~$< v
  <$~5~km\,s$^{-1}$}. The coefficients drop significantly for seven filtergrams,
i.e., the code apparently focusses on the large velocities. Using only the
inner three filtergrams therefore seems to be the most effective choice for
the reconstruction of velocities up to about 20~km\,s$^{-1}$. From this point
on, the velocities exceed the wavelength coverage of these filtergrams
and the outer filtergrams have to be employed. 

\begin{figure}[t]
 \centering
 \resizebox{0.99\hsize}{!}{\includegraphics[width=0.5\textwidth]{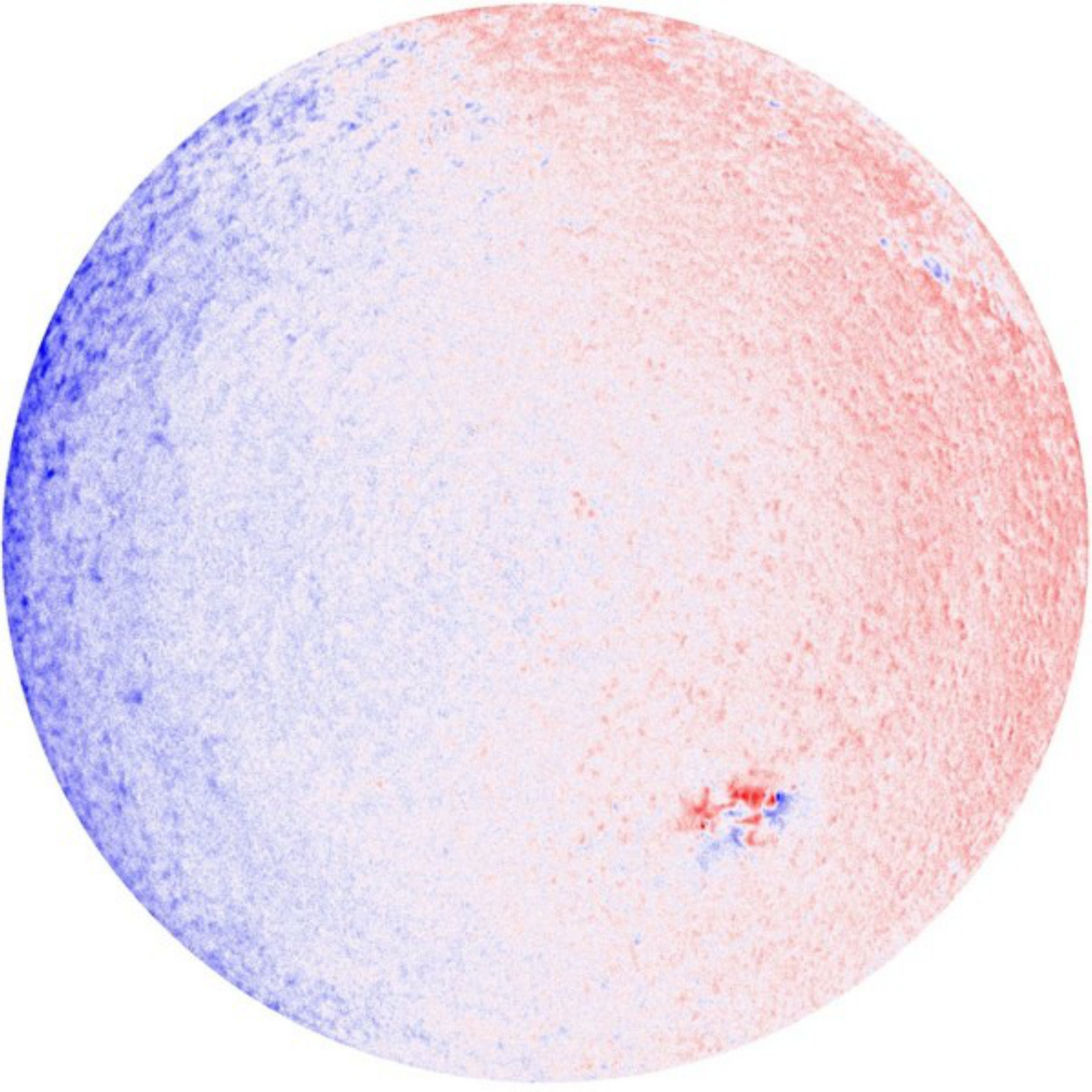}$\;\;$\includegraphics[height=0.37\textheight]{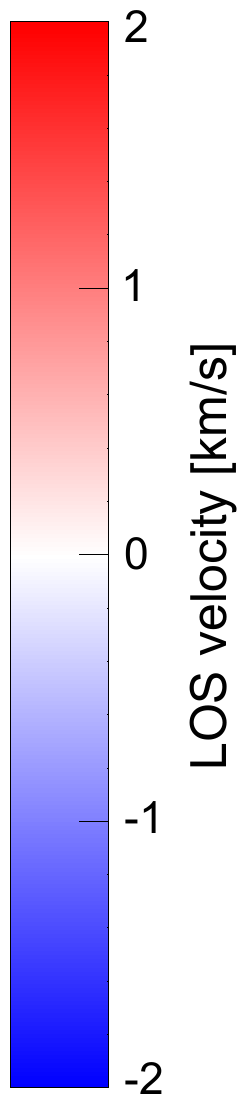}}
 \caption{\ion{He}{i} LOS velocity for the full disk on 6 July 2009, determined
   from averaged ChroTel data (UTC 15:58-16:39) using three filtergrams
   and the mean $\alpha_i$ parameters from all three calibration maps.} 
  \label{fig:he_fd_velo}
\end{figure}

An FWHM of 8.1~km\,s$^{-1}$ results in \mbox{$\,\sigma=\,$3.4~km\,s$^{-1}$}
for the use of three filtergrams. This however does not imply that
smaller velocities can not intrinsically be detected. We recall that a
lot of fine structure is less likely to be resolved with the lower spatial and spectal
resolution  of ChroTel, and that we instead observe a `time gradient' in the TIP-II scan, while
the ChroTel filtergrams represent a snapshot. In a full-disk
Dopplergram (see Fig.~\ref{fig:he_fd_velo}) created
with the mean $\alpha_i$ from several averaged ChroTel filtergrams,
the solar rotation is clearly visible. This means that the
accuracy of the LOS velocities must be more precise than
2~km\,s$^{-1}$ for small velocities on large scales.

\subsubsection{Different weightings} 
Two additional calibration runs were performed in attempting to improve the precision of
the velocity reconstruction. In the first run, only velocities of
\mbox{$-$5~$< v <$~5~km\,s$^{-1}$} were considered for the determination of the
$\alpha_i$ (``small velocities" in Table~\ref{table:pcc}). In the second
run, the fit results were weighted with the square of the absorption
depth in every pixel (``high abs. depth" in Table~\ref{table:pcc}). 

For the first run, the FWHM of the histogram for three filtergrams does
not change at all, and the correlation coefficients also vary only
very little compared to the regular run where all pixels were treated
the same. For the second run focussing on high absorption depths, the
FWHM drops insignificantly from 8.1 to 8.0~km\,s$^{-1}$,
and again the correlation coefficients display very little
variation. The different weightings therefore do not seem to have any
significant impact on the calibration quality.

\section{Discussion and conclusions}
The spatial resolution determined in the ChroTel images shows that they are at
least comparable in this respect to data from similar instruments such as
CHIP for \mbox{\ion{He}{i}}, PSPT for \mbox{\ion{Ca}{ii} K}, or the H$\alpha$
telescope of the Kanzelh\"ohe Observatory. The main advantage of
ChroTel however is the almost simultaneous acquisition of all three
spectral lines. Combining this information permits a more accurate estimation
of the temperatures and height profiles of structures.  

The data are taken with a cadence of three minutes and are also publicly
available for long-term \emph{in this cadence}, which is not the case
for most other full-disk instruments observing in these
lines. Although a cadence up to 30~s would be possible, three minutes were chosen as
a compromise between the quantity of data and also trying to catch
scarce and short-lived events in the regular mode of observation. The
adjustable observing mode nevertheless gives observers the opportunity
to take data at a much higher cadence if desired. 

The determination of LOS velocities from filtergrams in 
\mbox{\ion{He}{i}} offers an additional tool to study the dynamics of the
chromosphere. It has been shown that a precision of at least 
\mbox{3-4}~km\,s$^{-1}$ is possible with the method presented here, i.e., when
co-temporal spectrographic maps are available for calibration. Once
calibrated, the ChroTel Dopplergrams are a valuable supplement to
spectropolarimetric observations because they provide the
chromospheric velocity field in the full FOV with a high
cadence. This will be demonstrated in a future paper.

The inner five filtergrams seem to be sufficient for creating the
Dopplergrams, as the spread of the $\alpha_i$ and the correlation
coefficients for seven filtergrams indicate that the fits become
arbitrary. It might be valuable in the future to combine Dopplergrams
created with three and more filtergrams to optimize the results over 
the whole velocity range.

It is unclear to which extent the velocity calibration performed
within the scope of this work can be extrapolated and used for 
regions on the Sun other than the ones considered here. However,
that the solar rotation can be detected in full-disk Dopplergrams created
with the obtained calibration factors using data from another day 
provides confidence that these factors might be of a general nature and are not
only valid for the specific dataset used for the calibration. 

\begin{acknowledgements} 

ChroTel was built and put into operation with the help of many people
from the workshops of the participating institutes. Sincere thanks are
given to A.~Bernert, G.~Card, C.~Chambellan, H.P.~Doerr, R.~Friedlein,
R.~Hammer, T.~Hederer, T.~Keller, M.~Knobloch, A.~Lecinski, R.~Lull,
C.~Prahl, W.~Schmidt, M.~Sigwarth, T.~Sonner,  A.~Tritschler,
M.~Weissch\"adel, and O.~Wiloth.   

\noindent The ChroTel project was funded in part by the Deutsche
Forschungs\-gemeinschaft (DFG). 

\end{acknowledgements}

\bibliographystyle{aa} 
\bibliography{literature} 

\end{document}